# Analytical modeling of micelle growth. 2. Molecular thermodynamics of mixed aggregates and scission energy in wormlike micelles

Krassimir D. Danov [a], Peter A. Kralchevsky [a,*], Simeon D. Stoyanov [b,c,d], Joanne L. Cook [e], Ian P. Stott [e]

[a] *Department of Chemical and Pharmaceutical Engineering, Faculty of Chemistry and Pharmacy, Sofia University, Sofia 1164, Bulgaria*

[b] *Unilever Research & Development Vlaardingen, 3133AT Vlaardingen, The Netherlands*

[c] *Laboratory of Physical Chemistry and Colloid Science, Wageningen University, 6703 HB Wageningen, The Netherlands*

[d] *Department of Mechanical Engineering, University College London, WC1E 7JE, UK*

[e] *Unilever Research & Development Port Sunlight, Bebington CH63 3JW, UK*

ABSTRACT

*Hypotheses*

Quantitative molecular-thermodynamic theory of the growth of giant wormlike micelles in mixed nonionic surfactant solutions can be developed on the basis of a generalized model, which includes the classical "phase separation" and "mass action" models as special cases. The generalized model describes spherocylindrical micelles, which are simultaneously multicomponent and polydisperse in size.

*Theory*

The model is based on explicit analytical expressions for the four components of the free energy of mixed nonionic micelles: interfacial-tension, headgroup-steric, chain-conformation components and free energy of mixing. The radii of the cylindrical part and the spherical endcaps, as well as the chemical composition of the endcaps, are determined by minimization of the free energy.

*Findings*

In the case of multicomponent micelles, an additional term appears in the expression for the micelle growth parameter (scission free energy), which takes into account the fact that the micelle endcaps and cylindrical part have different compositions. The model accurately predicts the mean mass aggregation number of wormlike micelles in mixed nonionic surfactant solutions without using any adjustable parameters. The endcaps are enriched in the surfactant with smaller packing parameter that is better accommodated in regions of higher mean surface curvature. The model can be further extended to mixed solutions of nonionic, ionic and zwitterionic surfactants used in personal-care and house-hold detergency.

* Corresponding author. Tel.: +359 2 962 5310
  *E-mail address*: pk@lcpe.uni-sofia.bg (P.A. Kralchevsky)

## 1. Introduction

In the theory of micelle growth, two basic models have been developed, viz. the "phase separation" and "mass action" models [1-5]. The phase separation model is dealing with multicomponent but monodisperse micelles [4-11], whereas the mass action model describes polydisperse but single-component micelles [1-3, 12-21].

Experimentally, formation of large micellar aggregates is most frequently observed in mixed surfactant solutions, in which the micelles are simultaneously *multicomponent* and *polydisperse* in size [22-26]. Upon variation of solution's composition, peaks in viscosity have been often observed [27-31], which can be explained with the synergistic growth of giant entangled wormlike micelles and their transformations into disklike or multiconnected (branched) aggregates [32-37]. The prediction and control of micelle growth and formulation's viscosity are issues of primary importance for various practical applications [38-41].

Theoretically, molecular thermodynamic theories of micelle growth in mixed surfactant solutions were developed in studies by Ben-Shaul et al. [42-45], Nagarajan and Ruckenstein [46-48], and Blankschtein et al. [49-51]. In particular, on the basis of thermodynamic analysis of curvature effects, Gelbart et al. [42] were the first who pointed out that the compositions of the cylindrical part and the endcaps of a mixed spherocylindrical micelle (Fig. 1) should be, in general, different. Agreement between theory and experiment was achieved mostly with respect to the prediction of the critical micellization concentration. However, the quantitative prediction of the mean aggregation number of wormlike micelles and its dependence on micelle composition, temperature, surfactant chainlength, etc., remained a difficult problem; see Ref. [52], where a comprehensive review on wormlike micelles was recently published. Here, we focus our attention on the subject of the present article – achievement of agreement between theory and experiment with respect to the size of mixed wormlike micelles.

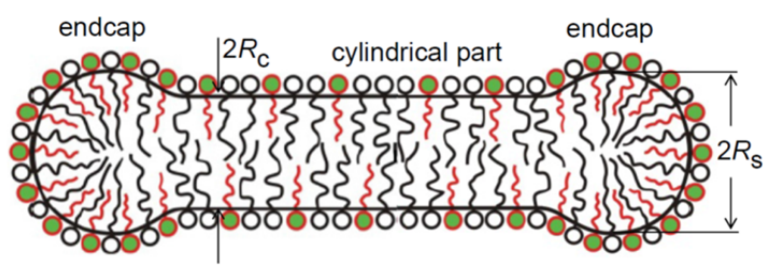

**Fig. 1.** Schematic presentation of a two-component spherocylindrical surfactant aggregate – wormlike (rodlike) micelle; $R_c$ and $R_s$ are the radii of the cylindrical part and the spherical endcaps.

To understand the difficulty of the aforementioned problem, let us consider the expression for the concentration dependence of the micelle mass average aggregation number [13,21,46,52]:

$$n_M = [K(X_S - X_S^o)]^{1/2}, \quad K = \exp E_{sc} \tag{1.1}$$

where $X_S$ is the total molar fraction of surfactant in the aqueous solution; $X_S^o$ is related to the intercept of the plot of $n_M^2$ vs. $X_S$, and $E_{sc} = \ln K$ is the *micelle growth parameter*. Eq. (1.1) is applicable to both single-component and multicomponent micelles (see Section 2).

For single-component *spherocylindrical* micelles, $E_{sc}$ can be expressed in the form [13,21,46,52]:

$$E_{sc} = n_s (f_s - f_c) / (k_B T) \tag{1.2}$$

where $n_s$ is the total aggregation number of the two micelle endcaps (with shapes of truncated spheres); $f_s$ and $f_c$ are the free energies per molecule in the endcaps and in the cylindrical part of the micelle, respectively; $k_B$ is the Boltzmann constant, and $T$ is the temperature. In other words, $E_{sc}k_B T$ is the *excess free energy* of the molecules in the spherical endcaps relative to the free energy of the same molecules if they were in the cylindrical part of the micelle. $E_{sc}k_B T$ represents also the micelle *scission free energy*, because the scission of a long wormlike micelle results in the appearance of two new endcaps [53]. Note that in the Cates' theory [32], the scission free energy is generally related to the average micellar length.

The enthalpy and entropy components of $E_{sc}$ have been determined by small-angle neutron scattering (SANS) and NMR measurements [54]. Theoretically, $E_{sc}$ was estimated using a potential of mean force [53], which was applied to simulations using the coarse grained dissipative particle dynamics (DPD) method [55]. In principle, the knowledge of the scission energy $E_{sc}$ is important also for kinetic models of relaxation of wormlike micelles [56,57] and for the rheological modelling of viscoelastic solutions containing giant micelles [32].

In Eqs. (1.1) and (1.2), typically $E_{sc}$ varies in the range 15–30 $k_B T$ units (see Section 2), $K$ – in the range $10^6$–$10^{13}$, $n_s$ – in the range 60–120, and $f_s - f_c$ varies in the range 0.125–0.50 $k_B T$. Hence, an inaccuracy of the order of 0.1 $k_B T$ in the calculation of $f_s - f_c$ would be strongly amplified when multiplied by (the relatively large) $n_s$ and then put in the argument of an exponential function to estimate $K$ and $n_M$; see Eqs. (1.1) and (1.2). In other words, the difference $f_s - f_c$ must be very accurately predicted by the theory in order to achieve a quantitative agreement with the experiment.

In Ref. [52], we developed a quantitative molecular-thermodynamic theory of $E_{sc}$ for single-component nonionic wormlike micelles. Analytical expression for $E_{sc}$ was derived, which presents $E_{sc}$ as a sum of three free-energy components related to interfacial tension, headgroup steric repulsion and chain conformations. The theory was verified against experimental data for the aggregation number $n_M$ of wormlike micelles from polyoxyethylene alkyl ethers, $C_nE_m$. The unknown temperature dependence of the excluded area per polyoxyethylene headgroup, $a_0(T)$, was determined from fits of experimental data with the theory. The agreement between theory and experiment was manifested through the fact that the values of $a_0(T)$ determined from independent sets of data for $C_nE_m$ surfactants with the same headgroup (but different chainlengths) collapsed on the same master curve.

As a next step toward a quantitative theory of wormlike micelles in *mixed* surfactant solutions, in Ref. [58] we extended the mean-field approach to the micelle chain-conformation free energy [52,59] to the case of two surfactants of different chainlengths. The derived analytical expressions for the chain-conformation components of $f_c$ and $f_s$ imply that the mixing of chains with different lengths in the micellar core is always nonideal and synergistic, and promotes micellization and micelle growth.

The goal of the present study is to extend the quantitative molecular-thermodynamic theory from Refs. [52,58] to the case of mixed nonionic wormlike micelles and to compare the theoretical predictions with experimental data. For this goal, in Section 2 we systematize available experimental data for binary mixtures of nonionic $C_nE_m$ surfactants to obtain values of $E_{sc}$ at different temperatures and micelle compositions. Next, in Section 3 the molecular thermodynamics of solutions containing multicomponent and polydisperse micelles is presented. In Section 4, the general thermodynamics is applied to the case of mixed spherocylindrical (wormlike) micelles (Fig. 1). It is shown that in the case of mixed micelles, Eq. (1.2) for $E_{sc}$ contains an additional term, which takes into account the fact that the micelle endcaps and the cylindrical part have different compositions. Section 5 is dedicated to the molecular aspects of the model – the analytical expressions for the four components of micelle free energy are generalized to the case of mixed micelles. Section 6 describes the procedure for numerical minimization of the analytical expression for free energy of a spherocylindrical micelle and numerical results are reported. Finally, Section 7 is dedicated to the comparison of theory and experiment for mixed wormlike micelles of nonionic surfactants. A serious challenge to the developed theory is that all physical parameters are known, so that there are no adjustable parameters. The theory takes this test successfully: excellent agreement with the experimental data is obtained without using any adjustable parameters.

## 2. Systematization of experimental data for binary mixtures of nonionic surfactants

Systematic light-scattering data for the growth of wormlike micelles in binary mixed solutions of nonionic surfactants, polyoxyethylene alkyl ethers, $C_nE_m$, at different temperatures have been published by Imanishi and Einaga [24]. Experimental results have been obtained for mixed solutions of two surfactants (i) with the same *alkyl chain*, $C_{14}E_5$ and $C_{14}E_7$, and (ii) with the same *polyoxyethylene chain*, $C_{14}E_5$ and $C_{10}E_5$. From the weight average molar mass of the micellar aggregates, $M_w$, determined by static light scattering [24], we calculated the micelle mass average aggregation number, $n_M = M_w / \bar{M}$, where, $\bar{M} = M_1 y_1 + M_2 y_2$ is the mean molar mass of the surfactant molecules; $M_1$ and $M_2$ are their molar masses; $y_1$ and $y_2$ are their molar fractions in the binary mixture ($y_1 + y_2 = 1$). The molar masses of the investigated surfactant molecules are $M_{C10E5}$ = 378.55 g/mol; $M_{C14E5}$ = 434.65 g/mol, and $M_{C14E7}$ = 522.76 g/mol.

As seen in Figs. 2 and 3, the data for $n_M$ from Ref. [24] are in excellent agreement with Eq. (1.1). The different panels correspond to different input weight fractions of $C_{14}E_5$ in the binary surfactant mixture,

$$w_j = \frac{m_j}{m_1 + m_2}, \quad j = 1, 2 \qquad (2.1)$$

where $m_1$ and $m_2$ are the masses of the two surfactants in the solution. Each straight line in Figs. 2 and 3 corresponds to a fixed temperature denoted in the figure. The data indicate that $n_M$ increases with the rise of both $X_S$ and $T$. The effect of $T$ can be explained with dehydration of the polyoxyethylene chains with the rise of temperature, which leads to enhanced intersegment attraction and compaction of the surfactant headgroups [52,60]. At the highest studied total surfactant concentrations, $n_M$ varies between ca. 2700 (for $T$ = 20 °C in Fig. 2a) to ca. 130,000 (for $T$ = 27 °C in Fig. 3c); in most cases, $n_M \sim 10^4$.

Table 1 summarizes the values of the dimensionless energy $E_{sc}$ (in $k_BT$ units) determined from the slopes of the experimental curves in Figs. 2 and 3 in accordance with Eq. (1.1). The values of $E_{sc}$ at $w_{C14E5}$ = 0 and 100 % are obtained from analogous plots in Ref. [52]. In Section 7, the values of $E_{sc}$ in Table 1 are used to test the theoretical model.

In the case of $C_{14}E_5 + C_{14}E_7$ (the same alkyl chains) at fixed $T$, both $E_{sc}$ and $n_M$ increase with the rise of $w_{C14E5}$, i.e. with the increase of the weight fraction of the surfactant with smaller headgroups. For example, at $T$ = 25 °C, we have $E_{sc}$ = 17.0, 22.5, 25.2 and 27.6 at $w_{C14E5}$ = 0, 25.1, 50 and 75 %, respectively. This fact is related to the circumstance that the decrease of the average area per headgroup with the rise of $w_{C14E5}$ favors the formation of bigger micelles of lower mean surface curvature.

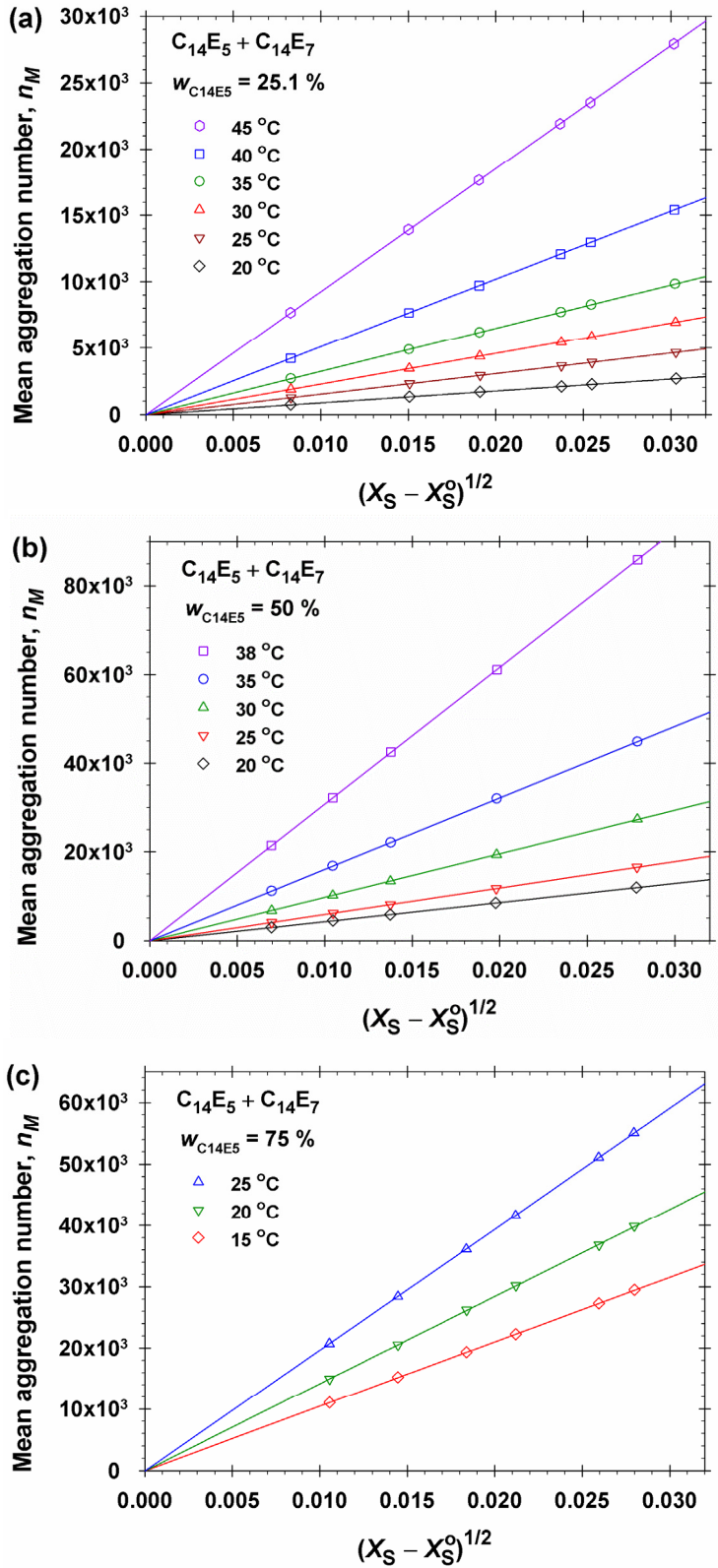

**Fig. 2.** Plots of the experimental micelle mean mass aggregation number, $n_M$, vs. $(X_S - X_S^o)^{1/2}$ in accordance with Eq. (1.1) for mixed micelles of $C_{14}E_5$ and $C_{14}E_7$ at various temperatures and at three different weight fractions of $C_{14}E_5$: (a) $w_{C14E5}$ = 25.1 %; (b) $w_{C14E5}$ = 50 %, and (c) $w_{C14E5}$ = 75 %. $X_S$ is the total surfactant molar fraction in the aqueous solution; $X_S^o$ is a constant parameter – see the text.

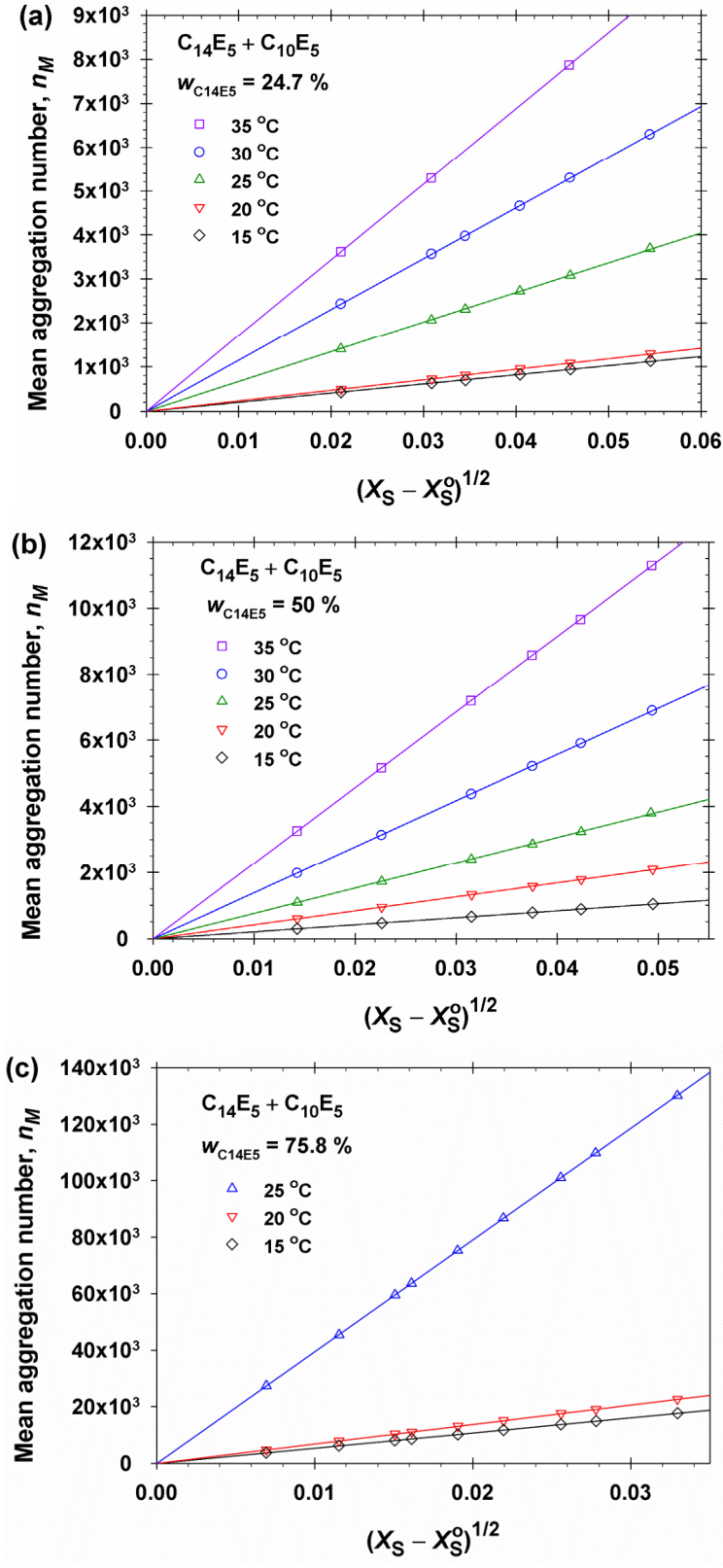

**Fig. 3.** Plots of the experimental micelle mean mass aggregation number, $n_M$, vs. $(X_S - X_S^o)^{1/2}$ in accordance with Eq. (1.1) for mixed micelles of $C_{14}E_5$ and $C_{10}E_5$ at various temperatures and at three different weight fractions of $C_{14}E_5$: (a) $w_{C14E5} = 24.7\ \%$; (b) $w_{C14E5} = 50\ \%$, and (c) $w_{C14E5} = 75.8\ \%$. $X_S$ is the total surfactant molar fraction in the aqueous solution; $X_S^o$ is a constant parameter – see the text.

**Table 1.** Micelle growth parameter (scission free energy in $k_BT$ units), $E_{sc}$, for wormlike micelles in mixed solutions of $C_{14}E_5 + C_{14}E_7$ and $C_{14}E_5 + C_{10}E_5$ at different temperatures and weight fractions, $w_{C14E5}$.

| $w_{C14E5}$ [%] | $T$ [°C] | $E_{sc}$ | $w_{C14E5}$ [%] | $T$ [°C] | $E_{sc}$ |
|---|---|---|---|---|---|
| $C_{14}E_5 + C_{14}E_7$ | | | $C_{14}E_5 + C_{10}E_5$ | | |
| 0.0 | 25 | 17.0 | 0.0 | 25 | 17.6 |
| 0.0 | 30 | 18.5 | 0.0 | 30 | 18.6 |
| 0.0 | 35 | 20.5 | 0.0 | 35 | 19.3 |
| 0.0 | 40 | 22.2 | 0.0 | 40 | 20.3 |
| 0.0 | 45 | 23.5 | 0.0 | 42 | 20.7 |
| 0.0 | 50 | 24.5 | 24.7 | 15 | 18.5 |
| 0.0 | 55 | 25.1 | 24.7 | 20 | 18.8 |
| 25.1 | 20 | 21.4 | 24.7 | 25 | 20.0 |
| 25.1 | 25 | 22.5 | 24.7 | 30 | 21.9 |
| 25.1 | 30 | 23.3 | 24.7 | 35 | 22.7 |
| 25.1 | 35 | 24.0 | 50.0 | 15 | 18.5 |
| 25.1 | 40 | 24.9 | 50.0 | 20 | 19.8 |
| 25.1 | 45 | 26.1 | 50.0 | 25 | 21.1 |
| 50.0 | 20 | 24.6 | 50.0 | 30 | 22.3 |
| 50.0 | 25 | 25.2 | 50.0 | 35 | 23.3 |
| 50.0 | 30 | 26.2 | 75.8 | 15 | 25.0 |
| 50.0 | 35 | 27.2 | 75.8 | 20 | 25.5 |
| 50.0 | 38 | 28.5 | 75.8 | 25 | 29.0 |
| 75.0 | 15 | 26.3 | 100 | 15 | 27.8 |
| 75.0 | 20 | 27.0 | 100 | 20 | 28.4 |
| 75.0 | 25 | 27.6 | – | – | – |

In the case of $C_{14}E_5$ and $C_{10}E_5$ (the same headgroups) at fixed $T$, both $E_{sc}$ and $n_M$ increase with the rise of the weight fraction, $w_{C14E5}$, of the surfactant with longer alkyl chain. For example, at $T = 25$ °C, we have $E_{sc} = 17.6$, 20.0, 21.1 and 29.0 at $w_{C14E5} = 0$, 24.7, 50 and 75.8 %, respectively. This behavior is related to the circumstance that the rise of $w_{C14E5}$ causes increase of the volume of the micelle hydrocarbon core, which leads to the formation of bigger micelles.

In Refs. [24,61], values of the free energy parameter $g_2$ are reported. This parameter is related to the dimensionless excess free energy $E_{sc}$ by the equation:

$$\frac{g_2}{k_\text{B}T} = E_\text{sc} - \ln \bar{M} \tag{2.2}$$

where $\bar{M} = M_1 y_1 + M_2 y_2$ (g/mol) is the mean molar mass of the surfactant molecules. The values of $g_2$ are with about 6–7 $k_\text{B}T$ smaller than those of $E_\text{sc}$.

Our goal in the rest of this paper is to develop a quantitative theoretical model that predicts the values of $E_\text{sc}$ and $n_M$ for mixed micelles from nonionic surfactants.

## 3. Molecular thermodynamics of mixed micellar solutions

### 3.1. Free energy of a multicomponent micellar surfactant solution

The free energy of a mixed solution of $m$ surfactants, which contains micellar aggregates, can be presented in the form:

$$G = N_\text{W} g_\text{W} + \sum_{j=1}^{m} N_{1,j} g_{1,j} + \sum_{k>1} N_k g_k(N_k, \mathbf{k}, s) \tag{3.1}$$

Here, $N_\text{W}$ is the number of solvent (e.g., water) molecules and $g_\text{W}$ is the free energy per solvent molecule; $N_{1,j}$ is the number of molecules of $j$th surfactant in the form of free monomers and $g_{1,j}$ is the free energy per monomer; $N_k$ is the number of micelles of aggregation number $k$, and $g_k$ is the free energy of such micellar aggregate. For brevity, $\mathbf{k}$ denotes the composition of a micelle that consists of $k$ surfactant molecules:

$$\mathbf{k} = (k_1, k_2, ..., k_m), \quad k = k_1 + k_2 + ... + k_m \tag{3.2}$$

where $k_j$ denotes the number of molecules from the $j$th component in the respective micelle; $m$ is the number of surface-active components. It is assumed that micelles of different aggregation number $k$ could have different composition $(k_1, k_2, ..., k_m)$, but the micelles with the same $k$ have the same composition. Finally, the argument $s$ in Eq. (3.1) denotes that $g_k$ depends on parameters, which characterize the shape of the micellar aggregate. For example, in the case of a cylindrical aggregate this is the radius of cylinder, $R_\text{c}$; in the case of a spherical endcap (Fig. 1) this is the endcap radius, $R_\text{s}$.

In Eq. (3.1) we have neglected the contribution from the interaction between the micellar aggregates in $G$. The established good agreement between the theory based on Eq. (3.1) and the experiment (Section 7) indicates that this approximation is reasonable in a wide range of concentrations.

Taking into account contributions from the entropy of mixing, we can present the free energies per molecule/aggregate in the form:

$$g_W(N_W) = \mu_W^o - k_B T + k_B T \ln(X_W) \tag{3.3}$$

$$g_{1,j}(N_{1,j}) = \mu_{1,j}^o - k_B T + k_B T \ln(X_{1,j}), \quad j = 1, 2, \ldots, m \tag{3.4}$$

$$g_k(N_k, \mathbf{k}, s) = g_k^o(\mathbf{k}, s) - k_B T + k_B T \ln(X_k) \quad \text{for } k > 1 \tag{3.5}$$

where $\mu_W^o$ and $\mu_{1,j}^o$ are molecular standard chemical potentials; $g_k^o(\mathbf{k}, s)$ is the standard free energy of a micellar aggregate composed of $k$ monomers, and the mole fractions are defined as follows:

$$X_W = \frac{N_W}{N_W + N_S}, \quad X_{1,j} = \frac{N_{1,j}}{N_W + N_S}, \quad X_k \equiv \frac{N_k}{N_W + N_S} \tag{3.6}$$

$$N_S = N_{S,1} + N_{S,2} + \ldots + N_{S,m} \tag{3.7}$$

$N_{S,j}$ is the number of surfactant molecules from the $j$th component in the solution, and $N_S$ is the total number of surfactant molecules.

The molecules of each surfactant component are distributed in a certain way between the aggregates of different size in the micellar solution. The mass conservation demands that their total number, $N_{S,j}$, must be constant:

$$N_{S,j} = N_{1,j} + \sum_{k>1} k_j N_k \tag{3.8}$$

In our subsequent analysis, we will use also the following definitions:

$$X_1 = \sum_{j=1}^{m} X_{1,j}, \quad X_S = \frac{N_S}{N_W + N_S} = X_1 + \sum_{k>1} k X_k \tag{3.9}$$

$X_1$ is the total molar fraction of the free surfactant monomers and $X_S$ is the total molar fraction of the input surfactant.

*3.2. Minimization of the free energy*

To find the equilibrium concentrations of all surfactant monomers and micellar aggregates in the solution, as well as the composition of the micelles, we have to minimize the free energy of the system with respect to the following variables: $N_{1,j}$ ($j = 1, \ldots, m$); $k_1, k_2, \ldots,$

$k_m$, and $N_k$ ($k >1$). At that, the *constraints* defined by Eqs. (3.2) and (3.8) have to be satisfied. For this reason, the Lagrange function, which has to be minimized, is:

$$G_L \equiv N_W g_W + \sum_{j=1}^{m} N_{1,j} g_{1,j} + \sum_{k>1} N_k g_k(\mathbf{k},s) + \sum_{j=1}^{m} \lambda_j (N_{S,j} - N_{1,j} - \sum_{k>1} k_j N_k) + \sum_{k>1} \xi_k (k - \sum_{j=1}^{m} k_j) \quad (3.10)$$

see Eq. (3.1). The variables $\lambda_j$ and $\xi_k$ are Lagrange multipliers. From a physical viewpoint, the equilibrium state should correspond to the minimum of $G_L$ with respect to all variables. Mathematically, to determine the values of all variables at the minimum of $G_L$, we have to set the first derivatives of $G_L$ with respect to these variables to be equal to zero. The conditions for minimum of $G_L$ with respect to $\xi_k$ and $\lambda_j$ give the constraints in Eqs. (3.2) and (3.8), as it should be. In view of Eq. (3.4), the minimization with respect to $N_{1,j}$ yields:

$$\lambda_j = \mu_{1,j}^o + k_B T \ln(X_{1,j}) \quad (j = 1, 2, ..., m) \quad (3.11)$$

Hence, at equilibrium $\lambda_j$ is equal to the chemical potential of the free monomers from the *j*th component. Furthermore, the minimization with respect to $N_k$ gives the relationship:

$$g_k^o(\mathbf{k},s) + k_B T \ln(X_k) = \sum_{j=1}^{m} k_j \lambda_j \quad (3.12)$$

Eq. (3.12) expresses the *mass action law* for a micelle of aggregation number $k$. Finally, the minimization with respect to $k_1, k_2, ..., k_m$ leads to:

$$\frac{\partial g_k}{\partial k_1} - \lambda_1 = \frac{\partial g_k}{\partial k_2} - \lambda_2 = ... = \frac{\partial g_k}{\partial k_m} - \lambda_m = \frac{\xi_k}{N_k} \quad (k > 1) \quad (3.13)$$

Note that the quantity $\mu_{k,j} = \partial g_k/\partial k_j$ is the chemical potential of a molecule from the *j*th component incorporated in a micelle of aggregation number $k$. Insofar as exchange of molecules between the micelles and monomers takes place, from a physical viewpoint we have to set zero the value of the variable $\xi_k$ at the minimum of $G_L$, i.e. at equilibrium $\xi_k = 0$. Then, Eq. (3.13) expresses the *equilibrium between micelles and monomers* with respect to all components.

Eq. (3.13), along with Eqs. (3.5) and (3.11), represents the basis of the "phase separation model" of micellization. In addition, Eq. (3.12) represents the basis of the "mass-action-law model"; see e.g. [1-3, 12-21]. Hence, the model presented here generalizes these two models in a natural way.

## 3.3. Micelle size distribution

Substituting $\lambda_j$ from Eq. (3.11) into Eq. (3.12) and taking inverse logarithm, we obtain:

$$X_k = X_{1,1}^{k_1} X_{1,2}^{k_2} ... X_{1,m}^{k_m} \exp[-\frac{g_k^o(\mathbf{k},s) - \Sigma_j k_j \mu_{1,j}^o}{k_B T}] \quad \text{for } k > 1 \tag{3.14}$$

Eq. (3.14) represents the micelle size distribution in a general form; see e.g. Ref. [21]. However, this form is not convenient for numerical calculations. Before the computations, it is necessary to transform Eq. (3.14) in a more convenient form. For this goal, let us introduce the variabes:

$$x_j \equiv \frac{X_{1,j}}{X_1} \quad (j = 1, 2, ..., m), \quad x_1 + x_2 + ... + x_m = 1 \tag{3.15}$$

$$y_j \equiv \frac{k_j}{k} \quad (j = 1, 2, ..., m), \quad y_1 + y_2 + ... + y_m = 1 \tag{3.16}$$

Here, $x_j$ denotes the mole fraction of free surfactant monomers from the $j$th component defined on *water-free* basis, whereas $y_j = y_j(k)$ is the mole fraction of surfactant molecules from the $j$th component in a micelle of aggregation number $k$; see also Eq. (3.2). Using the above definitions, we transform the pre-exponential factor in Eq. (3.14) as follows:

$$X_{1,1}^{k_1} X_{1,2}^{k_2} ... X_{1,m}^{k_m} = X_1^k \exp[k \sum_{j=1}^m y_j (\ln X_{1,j} - \ln X_1)] = X_1^k \exp(k \sum_{j=1}^m y_j \ln x_j) \tag{3.17}$$

Then, Eq. (3.14) acquires the form:

$$X_k = X_1^k \exp(-\frac{\Phi}{k_B T}) \tag{3.18}$$

where

$$\Phi \equiv k[f_k(\mathbf{k},s) - \sum_{j=1}^m y_j (\mu_{1,j}^o + k_B T \ln x_j)] \tag{3.19}$$

$$f_k(\mathbf{k},s) \equiv \frac{g_k^o(\mathbf{k},s)}{k} \tag{3.20}$$

$f_k(\mathbf{k},s)$ has the meaning of mean free energy per molecule in a mixed micelle.

The quantity $\Phi$ defined by Eq. (3.19) has the meaning of *free energy of a mixed micellar aggregate*. Indeed, if we minimize $\Phi$ with respect to the variables $k_1, k_2, ..., k_m$, along with the constraint in Eq. (3.2) (at fixed $N_{1,j}$, $j = 1, ..., m$), we obtain again the equilibrium relationships in Eq. (3.13). This means that the minimum value of $\Phi$ corresponds to a micellar aggregate, which is in equilibrium with the micellar solution with respect to the exchange of all surfactant components. Thus, starting with the free energy of the *whole* micellar solution [$G$ in Eq. (3.1)], we could continue our analysis with the minimization of the free energy of a *separate* micellar aggregate that is in equilibrium with the environment [$\Phi$ in Eq. (3.19)].

## 4. Molecular thermodynamics of spherocylindrical micelles

### 4.1. Minimization of the free energy of a micellar aggregate

The general equations derived in Section 3 can be applied to micelles of any specific shape, e.g. spherical, spheroidal, spherocylindrical, discoidal, etc. Here, our goal is to derive the equilibrium relationships for spherocylindrical (wormlike) micelles (Fig. 1) in the general case of different compositions and different radii of the cylindrical part and the spherical endcaps. For this goal, in the micellar free energy $\Phi$, defined by Eq. (3.19), we separate the contributions from the micelle cylindrical part and from the spherical endcaps:

$$\Phi = n_c f_c(\mathbf{y}_c, R_c) - n_c \sum_{j=1}^{m} y_{c,j}(\mu_{1,j}^o + k_B T \ln x_j) \\ + n_s f_s(n_s, \mathbf{y}_s, R_c) - n_s \sum_{j=1}^{m} y_{s,j}(\mu_{1,j}^o + k_B T \ln x_j) \quad (4.1)$$

where $n_c$ and $n_s$ denote the number of surfactant molecules contained in the cylindrical part and in the endcaps, respectively; for brevity, $\mathbf{y}_c$ and $\mathbf{y}_s$ denote the compositions of the cylindrical part and the endcaps:

$$\mathbf{y}_c = (y_{c,1}, y_{c,2}, ..., y_{c,m}), \quad y_{c,1} + y_{c,2} + ... + y_{c,m} = 1 \quad (4.2)$$

$$\mathbf{y}_s = (y_{s,1}, y_{s,2}, ..., y_{s,m}), \quad y_{s,1} + y_{s,2} + ... + y_{s,m} = 1 \quad (4.3)$$

Here and hereafter, the subscripts 'c' and 's' refer to the cylindrical part and the spherical endcaps, respectively. In Eq. (4.1) $f_c$ is independent of the total number of surfactant molecules, $n_c$, because the micelle is assumed to be sufficiently long so that the end effects are negligible. At known volume per surfactant tail, the endcap radius $R_s$ is determined if the endcap aggregation number, $n_s$, the endcap composition, $\mathbf{y}_s$, and the cylinder radius, $R_c$, are known. For this reason, in Eq. (4.1) $R_s$ is not given as an independent argument of $f_s$.

The equilibrium composition and size of a spherocylindrical micelle of given aggregation number, $k = n_c + n_s$, correspond to the minimum of micelle free energy at constraints defined by Eqs. (4.2) and (4.3). Hence, we have to minimize the following Lagrange function:

$$\Phi_L = n_s f_s(n_s, \mathbf{y}_s, R_c) - n_s \sum_{j=1}^{m} y_{s,j}(\mu_{1,j}^o + k_B T \ln x_j) + \xi_s(1 - \sum_{j=1}^{m} y_{s,j})$$

$$+ n_c f_c(\mathbf{y}_c, R_c) - n_c \sum_{j=1}^{m} y_{c,j}(\mu_{1,j}^o + k_B T \ln x_j) + \xi_c(1 - \sum_{j=1}^{m} y_{c,j}) \quad (4.4)$$

where $\xi_c$ and $\xi_s$ are Lagrangian multipliers.

The conditions for minimum of $\Phi_L$ with respect to $R_c$ at fixed composition leads to

$$n_s \frac{\partial f_s}{\partial R_c} + n_c \frac{\partial f_c}{\partial R_c} = 0 \quad (4.5)$$

For sufficiently long spherocylindrical micelles, we have $n_c \gg n_s$, so that the first term in Eq. (4.5) is negligible and we obtain:

$$\frac{\partial f_c}{\partial R_c} \approx 0 \quad \text{for } n_c \gg n_s \quad (4.6)$$

The condition for minimum of $\Phi_L$ with respect to the mole fractions $y_{s,j}$ and $y_{c,j}$ (at fixed $n_s$, $n_c$ and $R_c$) leads to:

$$\frac{\partial f_s(n_s, \mathbf{y}_s, R_c)}{\partial y_{s,j}} = \mu_{1,j}^o + k_B T \ln x_j + \frac{\xi_s}{n_s} \quad (j = 1, 2, ..., m) \quad (4.7)$$

$$\frac{\partial f_c(\mathbf{y}_c, R_c)}{\partial y_{c,j}} = \mu_{1,j}^o + k_B T \ln x_j + \frac{\xi_c}{n_c} \quad (j = 1, 2, ..., m) \quad (4.8)$$

Note that each of Eqs. (4.7) and (4.8) contains $m$ different equations, because the logic of the Lagrange minimization procedure based on Eq. (4.4) demands all $y_{s,j}$ and $y_{c,j}$ to be formally treated as independent variables, despite the constraints in Eqs. (4.2) and (4.3). This fact should be taken into account when calculating the partial derivatives in Eqs. (4.7) and (4.8); see Appendix A.

Having in mind that (by definition) the free energies of the endcaps and of the cylindrical part are $g_s = n_s f_s$ and $g_c = n_c f_c$, and that the derivatives in Eqs. (4.7) and (4.8) are taken at fixed $n_s$ and $n_c$, we obtain:

$$\frac{\partial f_s(n_s, \mathbf{y}_s, R_c)}{\partial y_{s,j}} = \frac{\partial g_s}{\partial n_{s,j}} = \mu_{s,j}, \quad \frac{\partial f_c(\mathbf{y}_c, R_c)}{\partial y_{c,j}} = \frac{\partial g_c}{\partial n_{c,j}} = \mu_{c,j}, \tag{4.9}$$

where $n_{s,j} = n_s y_{s,j}$ and $n_{c,j} = n_c y_{c,j}$ are the numbers of molecules in the spherical endcaps and in the cylindrical part of the $j$th component, and $\mu_{s,j}$ and $\mu_{c,j}$ are the respective chemical potentials. Hence, Eqs. (4.7) and (4.8) represent the conditions for *chemical equilibrium* between the endcaps and the cylindrical part of the micelle with the free surfactant monomers in the solution (as well as between the endcaps and the cylindrical part, themselves). Then, in analogy with Eq. (3.12) we have to set the equilibrium values of the Lagrangian multipliers to be equal to zero, viz. $\xi_c = \xi_s = 0$.

*4.2. Size distribution of the spherocylindrical micelles*

For sufficiently long micelles ($n_c \gg n_s$), the local properties in the cylindrical part of the micelle become independent on its total aggregation number, $k$. Because the spherical endcaps are in chemical equilibrium with the cylindrical part, their properties are also independent of $k$. Then, the micelle free energy, $\Phi$ in Eq. (4.1), becomes a linear function of $k$:

$$\Phi = Ck + E_{sc} k_B T \tag{4.10}$$

where the slope $C$ and the intercept $E_{sc} k_B T$ are defined as follows:

$$C \equiv f_c(\mathbf{y}_c, R_c) - \sum_{j=1}^{m} y_{c,j}(\mu^o_{1,j} + k_B T \ln x_j) \tag{4.11}$$

$$E_{sc} k_B T \equiv n_s[f_s(n_s, \mathbf{y}_s, R_c) - f_c(\mathbf{y}_c, R_c)] - n_s \sum_{j=1}^{m} (y_{s,j} - y_{c,j})(\mu^o_{1,j} + k_B T \ln x_j) \tag{4.12}$$

and the relation $n_c = k - n_s$ has been used. Note that $C$ is independent of the properties of the spherical endcaps, which are taken into account by $E_{sc}$. The quantity $E_{sc} k_B T / n_s$ represents the mean *excess free energy* per molecule in the spherical endcaps with respect to a molecule in the cylindrical part of the micelle. (Indeed, if we formally set $f_s = f_c$ and $y_{s,j} = y_{c,j}$, then Eq. (4.12) would give $E_{sc} = 0$.)

Physically, the breakage of a wormlike micelle to two parts is accompanied with the formation of two new endcaps. Hence, the excess free energy of the two endcaps, $E_{sc}k_BT$, can be identified with the reversible work for breakage of a long wormlike micelle, termed also *free energy of scission* [53,54].

In the special case of single-component micelles, $y_{s,j} = y_{c,j} = 1$ and Eq. (4.12) yields $E_{sc} = n_s(f_s - f_c)/(k_BT)$, which coincides with the definition for $E_{sc}$ in Ref. [52].

Substituting Eq. (4.10) in the micelle size distribution, Eq. (3.18), we obtain

$$X_k = \frac{q^k}{K}, \quad q \equiv \frac{X_1}{X_B}, \quad X_B \equiv \exp\left(\frac{C}{k_BT}\right) \tag{4.13}$$

$$K \equiv \exp(E_{sc}) \tag{4.14}$$

We recall that the size distribution defined by Eq. (4.13) holds for sufficiently large spherocylindrical micelles, for which $n_c \gg n_s$.

*4.3. Mean aggregation number by mass and by number*

By definition, the weight average molar mass of the micellar aggregates is:

$$M_w = \left(\sum_{k>1} M_{a,k}^2 N_k\right) / \left(\sum_{k>1} M_{a,k} N_k\right) \tag{4.15}$$

where $M_{a,k}$ is the mass of a micelle of aggregation number $k$:

$$M_{a,k} = k\sum_{j=1}^{m} M_j y_j(k) = k\overline{M}(k) \tag{4.16}$$

$$\overline{M}(k) = \sum_{j=1}^{m} M_j y_j(k) \tag{4.17}$$

$M_j$ is the molar mass of the $j$th surfactant component and $\overline{M}(k)$ is the mean molar mass for the molecules in a micelle of aggregation number $k$.

In the case of long spherocylindrical micelles, $n_c \gg n_s$, the micelle composition (with high precision) coincides with the composition of micelle cylindrical parts, i.e. $y_j \approx y_{c,j} =$ const. In other words, $y_j$ is independent of $k$. In view of Eq. (4.17), $\overline{M}$ also becomes independent of $k$. Then, the mean mass micelle aggregation number, $n_M$, can be expressed in the form:

$$n_M = \frac{M_w}{\overline{M}} = (\sum_{k>1} k^2 N_k)/(\sum_{k>1} k N_k) \tag{4.18}$$

see Eqs. (4.15) and (4.16). Finally, in view of Eqs. (3.6), (3.9) and (4.13) we obtain:

$$n_M = (\sum_{k>1} k^2 X_k)/(\sum_{k>1} k X_k) \approx 2[K(X_S - X_S^o)]^{1/2} \tag{4.19}$$

Likewise, the number-average micelle aggregation number, $n_N$, is

$$n_N = (\sum_{k>1} k X_k)/(\sum_{k>1} X_k) \approx [K(X_S - X_S^o)]^{1/2} \tag{4.20}$$

The derivation of the approximate expressions in the right-hand sides of Eqs. (4.19) and (4.20) can be found, e.g., in Refs. [21,52]. $X_S^o$ is proportional to the intercept of the plot of $n_M^2$ or $n_N^2$ vs. $X_S$. Typically, $X_S^o$ is of the same order of magnitude as $X_1$, but $X_S^o$ is not identical with $X_1$ because of a contribution from the smaller micelles, for which the linear dependence in Eq. (4.10) does not hold [52].

Eq. (4.19) is in excellent agreement with the experimental data for the mean mass aggregation number, $n_M$, for mixed micelles of nonionic surfactants; see Figs. 2 and 3. In view of Eqs. (4.14) and (4.19), the slope of each straight line in these figures is equal to $2\exp(E_{sc}/2)$. Thus, from the experimental slopes one can determine the value of $E_{sc}$ for the respective micellar solution and temperature (Table 1). Our next goal is to compare the experimental values of $E_{sc}$ with the theoretical $E_{sc}$ values predicted by the molecular-thermodynamic model.

*4.3. Expression for $E_{sc}$ in terms of interaction energies*

Eliminating $\mu_{1,j}^o + k_B T \ln x_j$ between Eqs. (4.8) and (4.12), we derive:

$$E_{sc} k_B T \equiv n_s[f_s(n_s, \mathbf{y}_s, R_c) - f_c(\mathbf{y}_c, R_c)] - n_s \sum_{j=1}^m (y_{s,j} - y_{c,j}) \frac{\partial f_c(\mathbf{y}_c, R_c)}{\partial y_{c,j}} \tag{4.21}$$

Formally, to obtain Eq. (4.21) it is not necessary to set the last term $\xi_c/n_c$ in Eq. (4.8) equal to zero; in view of the sums in Eqs. (4.2) and (4.3) it is sufficient that $\xi_c/n_c$ is independent of $j$.

The standard free energies per molecule in the micelle cylindrical part and spherical endcaps, $f_c$ and $f_s$, can be expressed in the form:

$$f_c = \sum_{j=1}^{m} \mu_{a,j}^o y_{c,j} + f_{c,\text{int}}, \quad f_s = \sum_{j=1}^{m} \mu_{a,j}^o y_{s,j} + f_{s,\text{int}} \tag{4.22}$$

where $\mu_{a,j}^o$ are standard chemical potentials of the surfactant molecules in the micellar aggregates, whereas the terms $f_{c,\text{int}}$ and $f_{s,\text{int}}$ take into account the interactions between the molecules in the respective parts of the micelle, including the free energy of mixing. The differentiation of the first term in Eq. (4.22) yields:

$$\frac{\partial f_c}{\partial y_{c,j}} = \mu_{a,j}^o + \frac{\partial f_{c,\text{int}}}{\partial y_{c,j}} \quad (j = 1, 2, ..., m) \tag{4.23}$$

In view of Eq. (4.23), the substitution of Eq. (4.22) in Eq. (4.21) leads to:

$$E_{sc} k_B T = n_s [f_{s,\text{int}}(R_s, \mathbf{y}_s, R_c) - f_{c,\text{int}}(\mathbf{y}_c, R_c)] - n_s \sum_{j=1}^{m} (y_{s,j} - y_{c,j}) \frac{\partial f_{c,\text{int}}(\mathbf{y}_c, R_c)}{\partial y_{c,j}} \tag{4.24}$$

In view of the relation between $R_s$ and $n_s$ (see Section 5.2), $R_s$ is chosen as an independent variable instead of $n_s$ in the argument of $f_{s,\text{int}}$. The last term in Eq. (4.24) is a collective contribution from all surface active species. It is related to the fact that the chemical compositions of the endcaps and the cylindrical part of the micelle are different. In the special case of single-component micelle ($m = 1$; $y_{s,1} = y_{c,1} = 1$), the last term is equal to zero and we arrive at the known result $E_{sc} k_B T = n_s(f_{s,\text{int}} - f_{c,\text{int}})$ [52].

It is important to note that in Eq. (4.24) there are no terms with $\mu_{a,j}^o$, that is $E_{sc}$ does not depend on the standard chemical potentials. [Eqs. (4.22) and (4.23) contain terms with $\mu_{a,j}^o$, but in Eq. (4.24) all of them have cancelled each other.] From a physical viewpoint, $E_{sc} k_B T$ represents the work for formation of two endcaps and, consequently, $E_{sc}$ is related to the change in energy due to the transfer of $n_s$ surfactant molecules from the cylindrical part to the endcaps, rather than to contributions from the internal molecular degrees of freedom, which are taken into account by $\mu_{a,j}^o$. For this reason, it is natural that $E_{sc}$ does not include contributions from $\mu_{a,j}^o$. In other words, to calculate theoretically the mean micelle aggregation numbers $n_M$ and $n_N$ one does not need the values of $\mu_{a,j}^o$; see Eqs. (4.14), (4.19), (4.20) and (4.24). The growth of long spherocylindrical micelles is controlled by the interactions between the molecules in the aggregates, which are taken into account by $f_{c,\text{int}}$ and $f_{s,\text{int}}$.

## 5. Molecular aspects of the model

*5.1. Formulation of the problem*

Our next goal is to calculate the interaction free energies (per molecule) $f_{c,\text{int}}$ and $f_{s,\text{int}}$ on the basis of information on the size and shape of the surface-active molecules incorporated in the micelle and the interactions between them. For mixed micelles composed of several nonionic surfactants, the interaction free energy can be expressed as a sum of four components:

$$f_{x,\text{int}} = f_{x,\text{mix}} + f_{x,\sigma} + f_{x,\text{hs}} + f_{x,\text{conf}}, \quad x = c, s \tag{5.1}$$

Here and hereafter, the subscript 'x' denotes quantities that refer to the cylindrical part of the micelle (x = c) or to the spherical endcaps (x = s). The first term in Eq. (5.1), $f_{x,\text{mix}}$, expresses the contribution of the free energy of ideal mixing, whereas the next three terms express contributions from interactions between the molecules in the micelle. In particular, $f_{x,\sigma}$ is the *interfacial tension* component, which takes into account the surface energy of contact of micelle hydrocarbon core with the outer aqueous phase; $f_{x,\text{hs}}$ is the *headgroup steric repulsion* component, which expresses the contribution from the repulsion between the surfactant headgroups on the micelle surface due to their finite size, and finally, $f_{x,\text{conf}}$ is the *chain-conformation* component, which expresses an energy contribution from the extension of surfactant chains in the micelles core with respect to their conformations in an ideal solvent; see illustrations in Ref. [52]. The last three components in Eq. (5.1) have been quantified in the special case of single-component micelles [52], but here we have to generalize the respective formulas for multicomponent micelles.

*Input parameters* used to calculate $f_{c,\text{int}}$ and $f_{s,\text{int}}$ are:

$$T, v_j, l_j, y_{c,j}, \text{ and } a_{0,j} \quad (j = 1, ..., m) \tag{5.2}$$

As usual, $T$ is the temperature; $v_j$ and $l_j$ are the volume and the length of the hydrocarbon tail of a molecule from the *j*th surface active component; $y_{c,1}, ..., y_{c,m}$ represent the composition of the cylindrical part of the micelles, which for long spherocylindrical micelles ($n_c \gg n_s$) is equal to the known input composition of the solution. Finally, $a_{0,j}(T)$ is the excluded area per headgroup of surfactant molecule of the *j*th component. The values of $a_{0,j}(T)$ for various polyoxyethylene alkyl ethers, $C_nE_m$, have been determined in Ref. [52]; see Eq. (6.5) and Table 3 therein. Formulas for calculation of $v_j$ and $l_j$, and other geometrical parameters are given in Section 5.2.

Alternatively, the cross-sectional area per headgroup, $a_{0,j}(T)$, can be determined theoretically, e.g., by using the Semenov mean field theory [59] to describe the conformations of the polyoxyethylene chains of the headgroups in water. At that, one should take into account the circumstance that with the rise of temperature the water undergoes a gradual transition from good solvent to poor solvent [60], which in a final reckoning leads to the appearance of cloud point for the nonionic surfactants. From this viewpoint, the theoretical prediction of $a_{0,j}(T)$ is a rather nontrivial task, which demands a separate study.

The parameters, which are to be determined by *minimization* of the free energy, are:

$$R_c, R_s, y_{s,1}, ..., y_{s,m} \tag{5.3}$$

As usual, $R_c$ and $R_s$ are the radii of the micelle hydrocarbon core in the cylindrical part and spherical endcaps, respectively; $y_{s,1}, ..., y_{s,m}$ is the composition of the spherical endcaps that are in chemical equilibrium with the cylindrical part of the micelle.

For example, in the case of *two-component* spherocylindrical micelles, $R_c$ is obtained by minimization of $f_{c,\text{int}}(R_c)$. Furthermore, with the obtained value of $R_c$ one determines $R_s$ and $y_{s,1}$ by minimization of $E_{sc}(R_c, R_s, y_{s,1})$ with respect to variations of $R_s$ and $y_{s,1}$ at given $R_c$; as usual, $y_{s,2} = 1 - y_{s,1}$; for details see Section 6.

*5.2. Molecular geometric parameters*

For a surfactant with alkyl chains of $n_j$ carbon atoms, the extended chainlength, $l_j$, and chain volume, $v_j$, can be calculated from the Tanford formulas [12]:

$$l_j = l(\text{CH}_3) + (n_j - 1)l(\text{CH}_2) \tag{5.4}$$

$$v_j = v(\text{CH}_3) + (n_j - 1)v(\text{CH}_2) \tag{5.5}$$

The volumes of the $\text{CH}_3$ and $\text{CH}_2$ groups, estimated from the temperature dependence of the volume of aliphatic hydrocarbons, are [46]:

$$v(\text{CH}_3) = [54.3 + 0.124(T - 298)] \times 10^{-3} \text{ nm}^3 \tag{5.6}$$

$$v(\text{CH}_2) = [26.9 + 0.0146(T - 298)] \times 10^{-3} \text{ nm}^3 \tag{5.7}$$

where $T$ is the absolute temperature. Temperature dependences of $l(\text{CH}_3)$ and $l(\text{CH}_2)$ have not been reported in the literature. We assume that these lengths are not sensitive to $T$ and use their values at 25 °C [12]:

$$l(CH_3) = 0.280 \text{ nm}, \quad l(CH_2) = 0.1265 \text{ nm} \tag{5.8}$$

The volume and surface area (of the hydrocarbon core) of the *cylindrical part* of the micelle are $V_c = \pi R_c^2 L$ and $A_c = 2\pi R_c L$, where $L$ is the length of cylinder. Hence,

$$A_c = \frac{2V_c}{R_c} = \frac{2}{R_c} n_c \sum_{j=1}^{m} y_{c,j} v_j \tag{5.9}$$

Then, the mean surface area per molecule in the cylindrical part of the micelle is:

$$a_c \equiv \frac{A_c}{n_c} = \frac{2}{R_c} \sum_{j=1}^{m} y_{c,j} v_j \tag{5.10}$$

Likewise, the volume and surface area (of the hydrocarbon core) of the *two spherical endcaps* (truncated spheres) are:

$$V_s = \frac{4}{3}\pi R_s^2 [R_s + (R_s^2 - R_c^2)^{1/2}] + \frac{2}{3}\pi R_c^2 (R_s^2 - R_c^2)^{1/2} \tag{5.11}$$

$$A_s = 4\pi R_s [R_s + (R_s^2 - R_c^2)^{1/2}] \tag{5.12}$$

Next, for a given composition, $y_{s,1}, \ldots, y_{s,m}$, one can calculate the number of molecules in the two spherical endcaps:

$$n_s = V_s / \sum_{j=1}^{m} y_{s,j} v_j \tag{5.13}$$

Furthermore, the mean surface area per molecule in the spherical endcaps, $a_s$, and their packing parameter, $p_s$, are:

$$a_s \equiv \frac{A_s}{n_s} = \frac{A_s}{V_s} \sum_{j=1}^{m} y_{s,j} v_j, \quad p_s = \frac{V_s}{A_s R_s} \tag{5.14}$$

The minimal value, $p_s = 1/3$, corresponds to hemispherical caps ($R_c = R_s$), whereas the maximal value, $p_s = 3/8$, corresponds to $R_s / R_c = 2/\sqrt{3} \approx 1.155$.

*5.3. Free energy of mixing*

For a mixture of molecules of different chainlength, the free energy of mixing (per molecule) can be expressed in the form [62]:

$$\frac{f_{x,\text{mix}}}{k_B T} = \sum_{j=1}^{m} y_{x,j} \ln \eta_{x,j}, \quad x = c, s \tag{5.15}$$

where $\eta_{x,j}$ is the volume fraction of the $j$-th surfactant chain in the micelle core:

$$\eta_{x,j} = \frac{y_{x,j} v_j}{\sum_{j=1}^{m} y_{x,j} v_j} \quad (j = 1, 2, ..., m) \tag{5.16}$$

Eq. (5.15) represents a generalization of the known expression from the Flory-Huggins theory [62].

*5.4. Interfacial tension component of the micelle free energy*

Generalizing the respective expression for single-component micelles [52] to the considered case of multicomponent micelles, we obtain:

$$f_{x,\sigma} = \sigma(a_x - a_0), \quad x = c, s \tag{5.17}$$

where $\sigma$ is the interfacial tension; $a_x$ is the surface area per molecule; see Eqs. (5.10) and (5.14), and $a_0$ is the mean surface area excluded by the surfactant headgroups of geometrical cross-sectional areas $a_{0,1}, a_{0,2}, ..., a_{0,m}$:

$$a_0 = \sum_{j=1}^{m} y_j a_{0,j} \tag{5.18}$$

With account for the micelle surface curvature, $\sigma$ is to be calculated from the Tolman formula [63,64]:

$$\sigma = \sigma_{ow}[1 + (\frac{1}{p_x} - 1)\frac{\delta_T}{R_x}]^{-1}, \quad x = c, s \tag{5.19}$$

Here, $p_c = 1/2$, $p_s$ is given by Eq. (5.14); $\sigma_{ow}$ is the interfacial tension between water and the mixed bulk oil phase, and $\delta_T$ is the Tolman length [52]:

$$\sigma_{ow} = 47.12 + 1.479(\sum_{j=1}^{m} y_{x,j} n_j)^{0.5422} - 0.0875(T - 293) \quad (\text{mN/m}) \tag{5.20}$$

$$\delta_T = 0.1456 \sum_{j=1}^{m} y_{x,j} l_j \quad (\text{nm}) \tag{5.21}$$

As before, $n_j$ ($j = 1, 2, ..., m$) is the number of the carbon atoms in the respective hydrocarbon chain.

*5.5. Headgroup steric repulsion component*

This component can be calculated by using the repulsion term in the two-dimensional van der Waals equation [52]:

$$f_{x,\text{hs}} = -k_B T \ln(1 - \frac{a_{\text{hs}}}{a_x}), \quad x = c,s \tag{5.22}$$

where $a_x$ is the surface area per molecule, see Eqs. (5.10) and (5.14), and $a_{\text{hs}}$ is the effective excluded area in the van der Waals model [65,66]:

$$a_{\text{hs}} = \sum_{i,j=1}^{m} y_{x,i} y_{x,j} a_{ij} \tag{5.23}$$

The diagonal elements of the matrix $a_{ij}$ are the respective headgroup cross-sectional areas, $a_{0,j}$:

$$a_{jj} = a_{0,j} \quad (j = 1, 2, ..., m) \tag{5.24}$$

The non-diagonal element, $a_{ij}$ ($i \neq j$), is identified with the area covered by a disk of radius equal to the arithmetic mean of the radii of the disks corresponding to components $i$ and $j$ [65,66]:

$$a_{ij} \equiv (\frac{a_{0,i}^{1/2} + a_{0,j}^{1/2}}{2})^2 \quad (i \neq j) \tag{5.25}$$

The validity of this model was proven in studies on the processing of surface tension isotherms of mixed surfactant solutions [66-68].

*5.6 Chain-conformation component of free energy*

*Surfactants of the same chainlength*, but of different headgroups. In this case, one can use the formula for identical chains of extended length $l$ [52]:

$$\frac{f_{x,\text{conf}}}{k_B T} = \frac{3\pi^2 R_x^2}{16 l_{\text{sg}} l} c_{\text{conf}}(p_x), \quad x = c,s \tag{5.26}$$

$$c_{\text{conf}}(p_x) = \frac{4 p_x^2}{1 + 3 p_x + 2 p_x^2} \tag{5.27}$$

where $l_{\text{sg}}$ is the length per segment in the chain. As suggested by Dill, Flory et al. [69,70], one can use the value $l_{\text{sg}} = 0.46$ nm, which is appropriate for alkyl chains. For the cylindrical part of the micelle, $p_c = 1/2$ and $c_{\text{conf}} = 1/3$, whereas for the spherical endcaps $p_s$ is given by Eq. (5.14).

The effect of headgroup size is taken into account by $f_{x,\sigma}$ and $f_{x,\text{hs}}$; see above. Because the equilibrium radii $R_c$ and $R_s$, and the endcap composition are obtained by minimization of the total interaction free energy, see Eq. (5.1), the headgroup sizes affect $f_{x,\text{conf}}$ through the equilibrium value of $R_x$ and $p_x$; see Eq. (5.26).

*Surfactants of different chainlengths.* In this case, expression for $f_{x,\text{conf}}$ is available only in the case of binary mixture of surfactants of extended chainlengths $l_1$ and $l_2$. By definition, it is assumed that $l_2 < l_1$. In this case, $f_{x,\text{conf}}$ can be calculated from the expression [58]:

$$\frac{f_{x,\text{conf}}}{k_B T} = \frac{3\pi^2 R_x^2}{16 l_{sg}} c_{\text{conf}}(p_x) \left[ \frac{y_{x,1}}{l_1} + \frac{y_{x,2}}{l_2} - \left(\frac{\bar{l}}{l_2^2} - \frac{\bar{l}}{l_1^2}\right) \beta_{\text{conf}} \right], \quad x = c,s \tag{5.28}$$

Here, $\bar{l} = y_{x,1} l_1 + y_{x,2} l_2$ is an average chainlength; $c_{\text{conf}}(p_x)$ is given by Eq. (5.27), and $\beta_{\text{conf}} = \beta_{\text{conf}}(p_x, \eta_{x,1})$ is the chain-conformation interaction parameter [58]:

$$\beta_{\text{conf}}(p_x, \eta_{x,1}) \equiv \frac{1}{c_{\text{conf}}(p_x)} \left\{ [b^2 - c_{\text{conf}}(p_x)] \eta_{x,1} + \frac{2}{p_x} \int_b^1 z(z^2 - b^2)^{1/2} (1-z)^{\frac{1-p_x}{p_x}} dz \right\} \tag{5.29}$$

where $\eta_{x,1}$ is the volume fraction of the chains of the surfactant of longer chainlength, and the parameter $b$ is defined as a solution of the equation [58]

$$\int_b^1 \frac{z(1-z)^{\frac{1-p_x}{p_x}}}{p_x (z^2 - b^2)^{1/2}} dz = \eta_{x,1} \quad (0 \le b \le 1); \quad x = c,s \tag{5.30}$$

for given values $p_x$ and $\eta_{x,1}$. In the case of *endcaps*, $b$ has to be calculated by numerical solution of Eq. (5.30), where $p_s$ is given by Eq. (5.14). In the case of *cylinder*, we have $p = p_c = 1/2$ and the integral in Eq. (5.30) can be taken analytically:

$$(1-b^2)^{1/2} - b^2 \ln\left[\frac{1}{b} + \left(\frac{1}{b^2} - 1\right)^{1/2}\right] = \eta_{c,1} \quad \text{(cylinder)} \tag{5.31}$$

Eq. (5.31) is a transcendental equation for $b$, which is to be solved numerically. As demonstrated in Ref. [58], the solution of Eqs. (5.30) and (5.31) for $b$ always exists ($0 \le b \le 1$). At that, $b$ is a monotonically decreasing function of $\eta_{x,1}$.

Physically, $b$ is the boundary between the outer and inner regions in the micelle interior. In the *outer* region, $0 < r/R_x < b$, the ends of the *shorter* chains are located; in the *inner* region, $b < r/R_x < 1$, the ends of the *longer* chains are located. Here, $r$ is a radial coordinate, with $r = 0$ at the surface of micelle hydrocarbon core and $r = R_x$ in the micelle center [58].

In the case of *endcaps*, the integral in Eq. (5.29) has to be calculated numerically with $p_x = p_s$ given by Eq. (5.14). In the case of *cylinder*, we have $p_x = p_c = 1/2$ and the integral in Eq. (5.29) can be taken analytically, which leads to the following expression for $\beta_{\text{conf}}$ [58]:

$$\beta_{\text{conf}} = (1-b^2)^{3/2} + \frac{3}{2} b^2 \eta_{c,1} - \eta_{c,1} \quad \text{(cylinder)} \tag{5.32}$$

For $0 \leq \eta_{x,1} \leq 1$, $\beta_{conf}$ vs. $\eta_{x,1}$ is a curve with maximum [58], which is zero at the endpoints: $\beta_{conf}(\eta_{x,1}=0) = \beta_{conf}(\eta_{x,1}=1) = 0$. In other words, for mixed micelles ($0 < \eta_{x,1} < 1$) the interaction parameter is always positive, $\beta_{conf} > 0$. Thus, Eq. (5.28) implies that the mixing of two surfactant with different chainlengths is always synergistic with respect to the chain conformation free energy, $f_{x,conf}$.

## 6. Minimization of the free energy of a spherocylindrical micelle

### 6.1. Thermodynamic and computational aspects

As already mentioned, the radius $R_c$ of micelle *cylindrical part* is determined by minimization of the free energy $f_c$ with respect to variations of $R_c$. Analogously, the radius $R_s$ of micelle *spherical endcaps* and their composition are determined by minimization of the excess free energy $E_{sc}$ with respect to variations of the endcap radius $R_s$ and the molar fractions $y_{s,1},\ldots,y_{s,m}$. In the considered case of long spherocylindrical micelles ($n_c \gg n_s$), the composition of micelle cylindrical parts, $y_{c,1},\ldots,y_{c,m}$, is fixed and determined by the input concentrations of surfactants.

In view of Eqs. (4.6), (4.22) and (5.1), the condition for minimum of $f_c$ with respect to $R_c$ can be presented in the form:

$$0 = \frac{\partial f_c}{\partial R_c} = \frac{\partial}{\partial R_c}\left(f_{c,mix} + f_{c,\sigma} + f_{c,hs} + f_{c,conf}\right) \qquad (6.1)$$

Physically, Eq. (6.1) expresses a condition for mechanical equilibrium of the cylindrical micelle. If such a local minimum of $f_c$ does not exist in the interval $0 < R_c \leq l_{max}$, where $l_{max}$ is the chainlength of the longest surfactant molecule, then equilibrium spherocylindrical micelles could not exist.

Explicit analytical expressions for the derivatives of the components of $f_c$ in the right-hand side of Eq. (6.1) can be found in Appendix A. Thus, Eq. (6.1) is transformed into an algebraic equation for $R_c$, which has been solved numerically.

The condition for minimum of $E_{sc}$ with respect to the endcap composition, $y_{s,1},\ldots,y_{s,m}$, is equivalent to chemical equilibrium between the endcaps and the micelle cylindrical part with respect to exchange of molecules of all surface active components. Indeed, in accordance with Eq. (4.21) at $T$ = const. $E_{sc}$ is a function of $2m + 2$ independent variables, viz. $E_{sc} = E_{sc}(n_s,\mathbf{y}_s,R_c,\mathbf{y}_c)$. By differentiation of Eq. (4.21) with respect to $y_{s,j}$ and setting the derivative of $E_{sc}$ equal to zero as a necessary condition for minimum, we obtain:

$$0 = k_B T \frac{\partial E_{sc}}{\partial y_{s,j}} = n_s \frac{\partial f_s(n_s, \mathbf{y}_s, R_c)}{\partial y_{s,j}} - n_s \frac{\partial f_c(\mathbf{y}_c, R_c)}{\partial y_{c,j}} = n_s(\mu_{s,j} - \mu_{c,j}) \quad (6.2)$$

At the last step, Eq. (4.9) has been used. Thus, we obtain

$$\mu_{s,j} = \mu_{c,j}, \quad j = 1,...,m \quad (6.3)$$

i.e., the minimum of $E_{sc}$ corresponds to *chemical equilibrium* between the endcaps and the cylindrical part of the micelle. This result once again indicates the importance of the last term in Eq. (4.21) – without this term the chemical equilibrium relation, Eq. (6.3), cannot be obtained.

Because the experimental data in Section 2 refer to binary surfactant mixtures, in our computations we minimized numerically the function $E_{sc}(R_s, y_{s,1})$, as given by Eq. (4.24), with respect to variations of $R_s$ and $y_{s,1}$. In view of Eq. (5.1), the derivative in Eq. (4.24) can be presented in the form:

$$\frac{\partial f_{c,int}}{\partial y_{c,1}} = \frac{\partial f_{c,mix}}{\partial y_{c,1}} + \frac{\partial f_{c,\sigma}}{\partial y_{c,1}} + \frac{\partial f_{c,hs}}{\partial y_{c,1}} + \frac{\partial f_{c,conf}}{\partial y_{c,1}} \quad (6.4)$$

The four derivatives in the right-hand side of Eq. (6.4) have been calculated analytically – see the respective expressions in Appendix A. By using these expressions, one can avoid numerical differentiation. The computations indicate that the last term in Eq. (4.24), which takes into account the different composition of the endcaps (relative to the micelle cylindrical part), is comparable with the other terms in Eq. (4.24) and is never negligible.

*6.2. Numerical results and discussion*

The input parameters are those in Eq. (5.2), where $l_j$ and $v_j$ are calculated from Eqs. (5.4)–(5.8). Further, $f_{c,int}$ and $f_{s,int}$ are calculated from Eq. (5.1), where the four free energy components are computed using equations and parameter values given in Sections 5.3–5.6.

As an illustration, Fig. 4a shows plots of $f_c$ vs. $R_c$ for the mixed micelles of $C_{14}E_5$ and $C_{14}E_7$ at three different compositions, $w_{C14E5} = 25.1$, 50 and 75 %, for which experimental data are presented in Table 1. The parameter values correspond to $T = 30$ °C, and $C_{14}E_5$ is chosen as component 1. In addition, Fig. 4b shows a contour plot of the function $E_{sc}(R_s, y_{s,1})$ for $w_{C14E5} = 50$ %, which corresponds to $y_{c,1} = 0.546$. The values of $E_{sc}$ are given at the respective isolines.

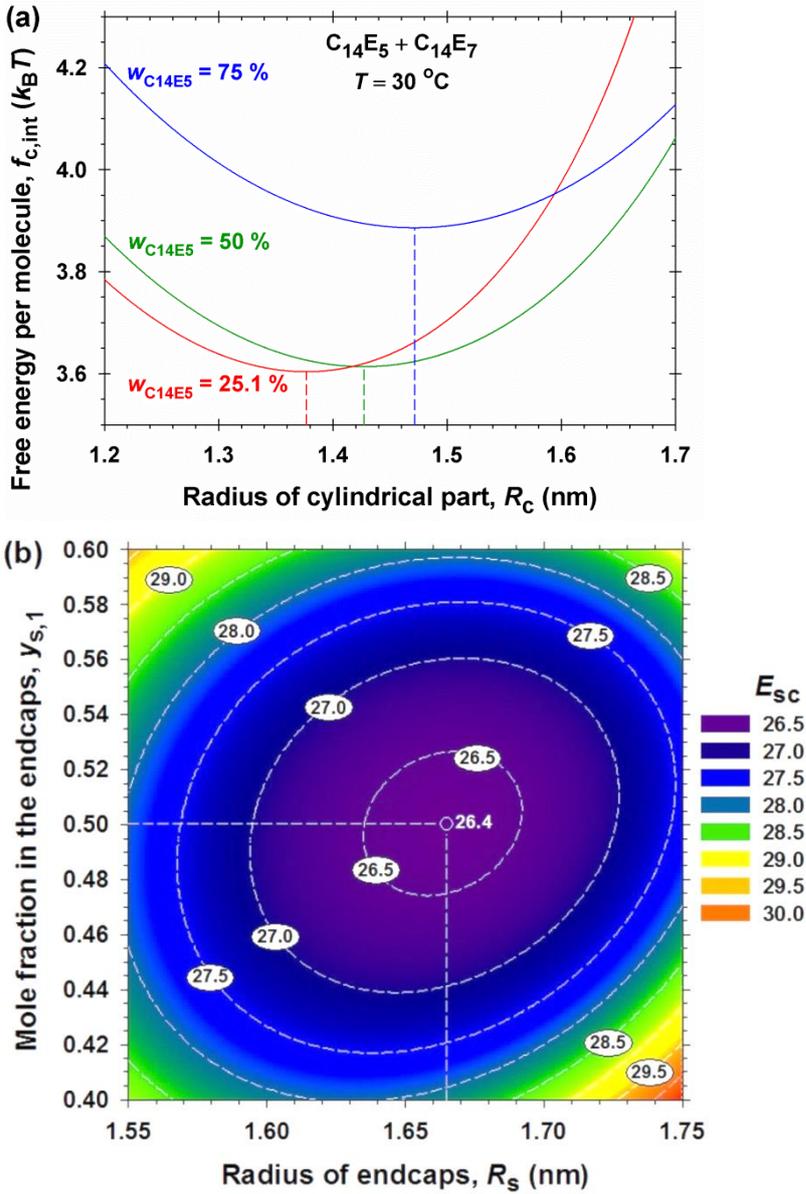

**Fig. 4**. Free energy minimization for mixed micelles from $C_{14}E_5 + C_{14}E_7$ at 30 °C. (a) Plot of $f_{c,int}$ vs. $R_c$ for the micelle cylindrical part at three different weight fractions of $C_{14}E_5$, $w_{C14E5}$, denoted in the figure. (b) Contour plot of the endcap excess free energy, $E_{sc} = E_{sc}(R_s,y_{s,1})$, at $w_{C14E5} = 50$ %. The equilibrium values of $R_c$, $R_s$ and $y_{s,1}$ are determined by the positions of the respective minima, which are shown by dashed lines.

Likewise, Fig. 5a shows plots of $f_c$ vs. $R_c$ for the mixed micelles of $C_{14}E_5$ and $C_{10}E_5$ at three different compositions, $w_{C14E5} = 24.7$, 50 and 75.8 %, for which experimental data are presented in Table 1. The parameter values correspond to $T = 25$ °C, and $C_{14}E_5$ is chosen as component 1. In addition, Fig. 5b shows a contour plot of the function $E_{sc}(R_s,y_{s,1})$ for $w_{C14E5} = 50$ %, which corresponds to $y_{c,1} = 0.466$.

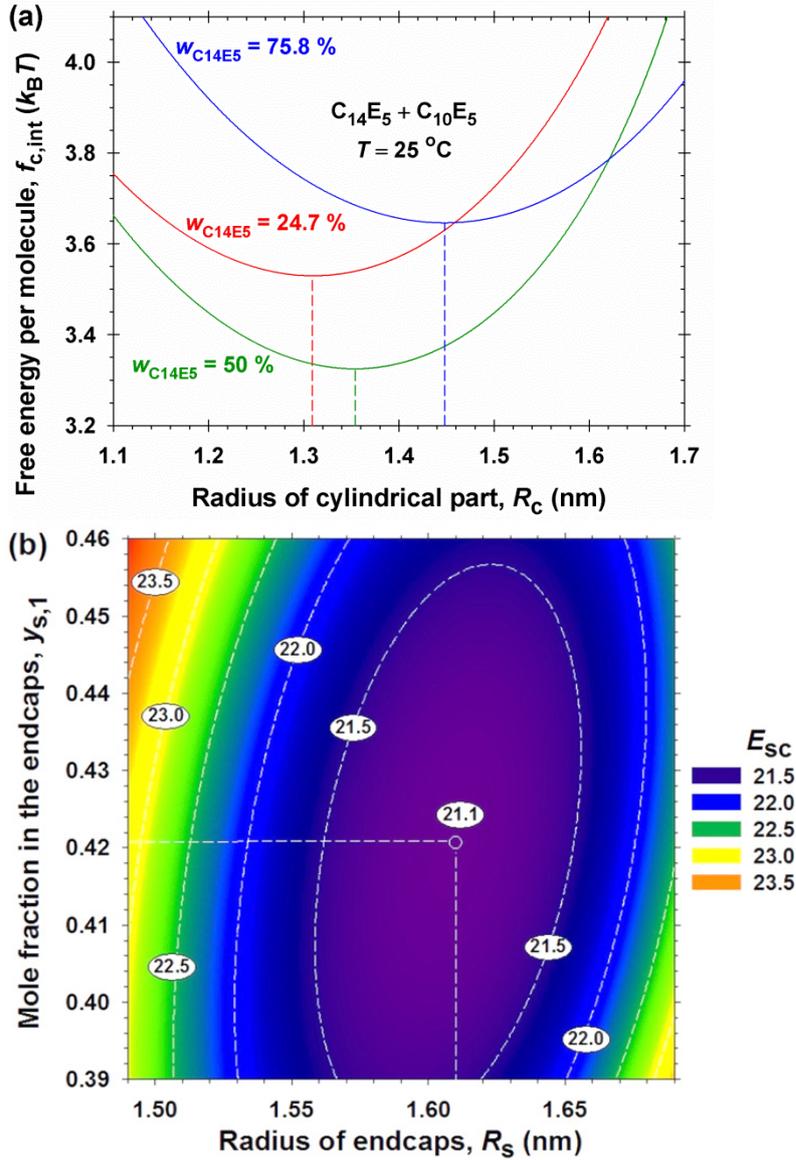

**Fig. 5**. Free energy minimization for mixed micelles from $C_{14}E_5 + C_{10}E_7$ at 25 °C. (a) Plot of $f_{c,\text{int}}$ vs. $R_c$ for the micelle cylindrical part at three different weight fractions of $C_{14}E_5$, $w_{C14E5}$, denoted in the figure. (b) Contour plot of the endcap excess free energy, $E_{sc} = E_{sc}(R_s, y_{s,1})$, at $w_{C14E5} = 50\ \%$. The equilibrium values of $R_c$, $R_s$ and $y_{s,1}$ are determined by the positions of the respective minima, which are shown by dashed lines.

All plots in Figs. 4 and 5 possess minima, which mean that mechanically equilibrium micelles exist and that their endcaps and cylindrical parts coexist in chemical equilibrium. The equilibrium values of $R_c$, $R_s$ and $y_{s,1}$ are those corresponding to the minima. In particular, in all cases the equilibrium values of $R_c$ and $R_s$ satisfy the physical requirement $0 < R_x \leq l_1$ (x = c,s), where $l_1 = 1.92$ nm for the longer $C_{14}$-alkyl chain; see Eqs. (5.4) and (5.8).

Note that the values $E_{sc} = 26.4$ and 21.1 at the minima of Figs. 4b and 5b practically coincide with the respective experimental values in Table 1, $E_{sc} = 26.2$ and 21.1. This is a remarkable coincidence, having in mind that no adjustable parameters have been used. As demonstrated in Section 7, such good agreement between theory and experiment is present for all other investigated compositions and temperatures.

## 7. Comparison of theory and experiment

*7.1. Experimental vs. theoretical values of $E_{sc}$*

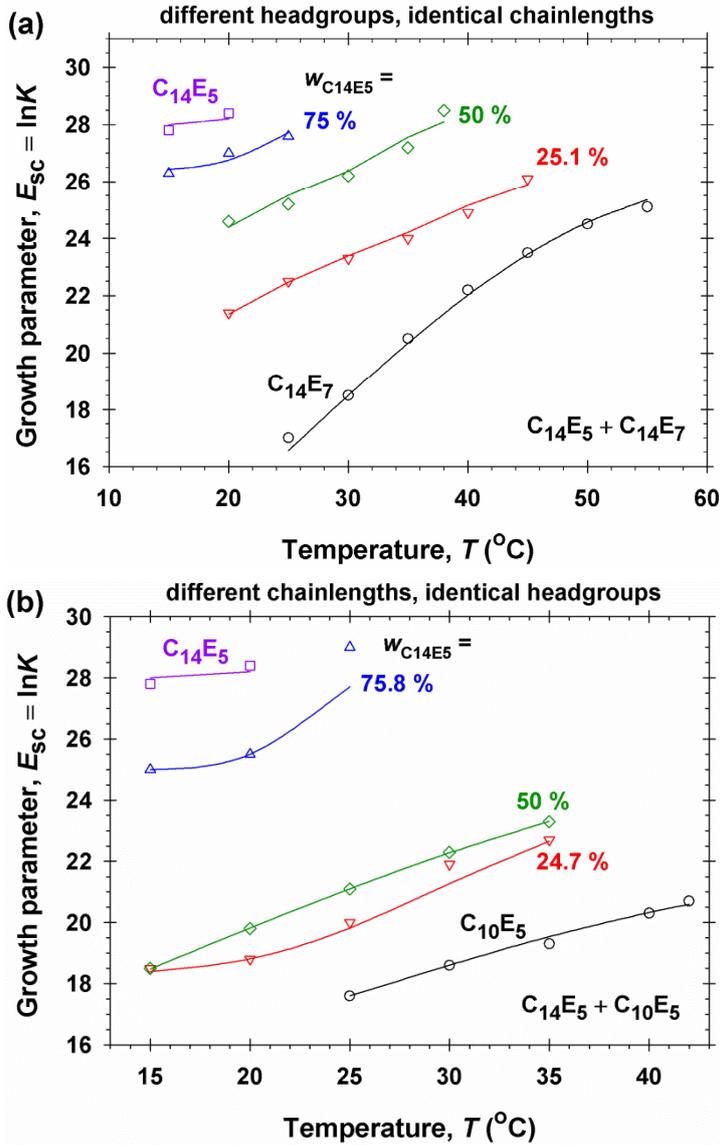

**Fig. 6.** Comparison of theory (solid lines) and experiment (points – data from Table 1) for mixed surfactant micelles at different weight fractions of $C_{14}E_5$, $w_{C14E5}$, denoted in the figure. Plots of the micelle growth parameter (scission energy in $k_BT$ units), $E_{sc}$, vs. temperature, $T$: (a) $C_{14}E_5$ and $C_{14}E_7$ (different headgroups) and (b) $C_{14}E_5$ and $C_{10}E_5$ (different chainlengths).

In Fig. 6a and b, the points are the experimental data for the micelle growth parameter $E_{sc}$ vs. temperature $T$ from Table 1, whereas the solid lines represent the predictions of theory for the respective composition of the surfactant mixture denoted in the figure. As already mentioned, the headgroup areas, $a_{0,j}(T)$, have been determined in Ref. [52] from fits of data for the growth of single-surfactant micelles. In the present study, the theoretical curves for mixed micelles are drawn without using any adjustable parameters. For both investigated systems, $C_{14}E_5+C_{14}E_7$ (identical alkyl chains) and $C_{14}E_5+C_{10}E_5$ (identical headgroups), there is an excellent agreement between theory an experiment, which confirms the adequacy of the developed theoretical model.

As explained in Section 6.2 (Figs. 4b and 5b), the values of $E_{sc}$ in Fig. 6 correspond to the minimum of the function $E_{sc}(R_s,y_{s,1})$. In our computations, the values of $E_{sc}$, $R_s$ and $y_{s,1}$ at the minimum were determined within an accuracy of three significant digits. In most cases, this accuracy was sufficient, but in isolated cases (as the curve for 50 % in Fig. 6a) the limited computational accuracy has led to small undulations in the calculated theoretical curve.

A comparison of the curves in Figs. 6a and b shows that in the case of different alkyl chains (Fig. 6b) the curves corresponding to almost equidistant $w_{C14E5}$ values are far from being equidistant (which is not the case in Fig. 6a). Thus, the curves for $w_{C14E5} = 24.7$ and 50 % are very close to each other, whereas the curve for $w_{C14E5} = 75.8$ % is situated far from them (Fig. 6b). Thus irregular behavior is related to the strong deviations from ideal mixing in the case of different alkyl chains, as discussed in Ref. [58], where the theory of the chain conformation free energy, $f_{x,conf}$, has been developed. The deviations from ideality are taken into account by the chain-conformation interaction parameter, $\beta_{conf}$ in Eq. (5.29), which is a non-monotonic function of the composition, $\eta_{x,1}$.

*7.2. Predicted values of the micellar parameters*

The theoretical calculation of the micelle growth parameter (scission energy) $E_{sc}$ includes calculation also of many other micellar parameters, such as the radii of the cylindrical part and of the endcaps, $R_c$ and $R_s$, as well as the aggregation number and composition of the endcaps, $n_s$, $y_{s,1}$ and $y_{s,2}$. It is difficult to directly measure the latter parameters of the mixed wormlike micelles, but the quantitative theory gives information for their values and variations.

In Fig. 7a and b, we compare the calculated plots of $R_c$ and $R_s$ vs. $y_{c,1}$ for the two investigated systems, $C_{14}E_5+C_{14}E_7$ and $C_{14}E_5+C_{10}E_5$, where $C_{14}E_5$ has been chosen as component 1. For both systems, $R_c < R_s$ (as it should be expected). The values of $R_c$ and $R_s$ satisfy the physical requirement $0 < R_x \leq l_1$ (x = c,s), where $l_1 = 1.92$ nm for the longer $C_{14}$-alkyl chain.

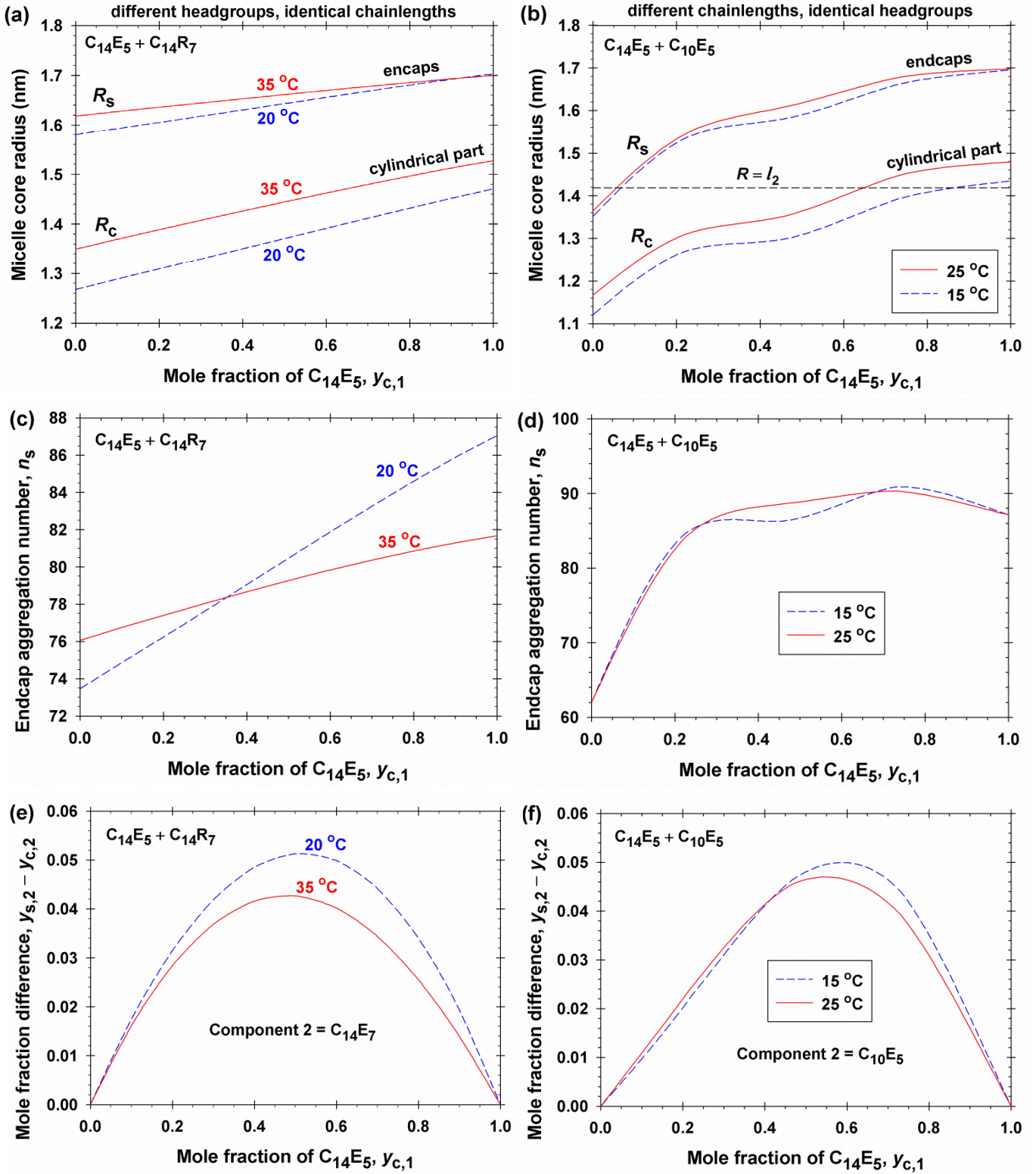

**Fig. 7.** Parameters of the mixed micelles predicted by the theory vs. the input mole fraction of $C_{14}E_5$, $y_{c,1}$, for the two investigated systems, $C_{14}E_5+C_{14}E_7$ (left) and $C_{14}E_5+C_{10}E_5$ (right). (a) and (b) Equilibrium radii of the cylindrical part and spherical endcaps, $R_c$ and $R_s$; (c) and (d) the aggregation number of the two endcaps together, $n_s$; (e) and (f) difference between the mole fractions of component 2 in the spherical endcaps and in the cylindrical part, $y_{s,2} - y_{c,2} = -(y_{s,1} - y_{c,1})$.

Note, however, that $R_c$, and especially $R_s$, can be essentially greater than the length of the shorter chain, $l_2$ (Fig. 7b). Both $R_c$ and $R_s$ increase with the rise of the input molar fraction of $C_{14}E_5$, $y_{c,1}$, which is the surfactant of smaller headgroup in Fig. 7a, but is the surfactant of longer tail in Fig. 7b. At that, the rise of $R_c$ and $R_s$ with $y_{c,1}$ in Fig. 7a is practically linear, whereas significant deviations from linearity are seen in Fig. 7b, which can be explained with the aforementioned nonideal mixing of chains of different length [58].

In Fig. 7c and d, we compare the calculated plots of the endcap aggregation number, $n_s$, vs. $y_{c,1}$ for the two investigated systems. Again, the plots for surfactants of identical chains (Fig. 7c) are practically linear, whereas those for surfactants of different chains show marked deviations from linearity, which are due to the nonideal chain mixing (see above). It is important to note that $n_s$ enters the expression for $E_{sc}$, Eq. (4.24), as a multiplier, and in turns, $E_{sc}$ enters the expression for micelle aggregation number $n_M$ in the argument of an exponent; see Eqs. (4.14) and (4.19). For this reason, the increase of $n_s$ is one of the main reasons for micelle growth. In Fig. 7c (at $T = 20$ °C), $n_s$ increases from 73.5 to 87, i.e. with ca. 18 %, whereas in Fig. 7d $n_s$ increases from 62 to 90, i.e. with ca. 45 %. The latter fact correlates with the circumstance that the biggest micelles of $n_M \approx 130{,}000$ are observed with the system $C_{14}E_5+C_{10}E_5$; see Fig. 3c.

In general, the endcaps have a composition, which is different from that of the cylindrical part of the micelle. The magnitude of this effect is illustrated in Figs. 7e and f, where the difference $y_{s,2} - y_{c,2}$ is plotted vs. the input molar fraction of $C_{14}E_5$, $y_{c,1}$. The results show that the endcaps are enriched in component 2, which is the component with smaller value of the packing parameter, $p_j = v_j/(a_{0,j} l_j)$, that is better accommodated in aggregate of higher surface curvature. In both Figs. 7e and f, the effect is the greatest at intermediate molar fractions, $y_{c,1} = 0.5 - 0.6$, and has a similar magnitude, $4.3 - 5.2$ %. Despite the relatively small values of $y_{s,j} - y_{c,j}$, the last term in Eq. (4.24) is comparable by magnitude with the other terms in this equation.

*7.3. Theory vs. experiment for $C_nE_m$ + n-dodecanol*

The data by Miyake and Einaga [61] on the growth of wormlike micelles in mixed solutions of $C_{10}E_5$ + n-dodecanol and $C_{12}E_6$ + n-dodecanol represent another set of experimental results, which allow verification of our theoretical model. The data in Ref. [61], which are originally presented in terms of $g_2$, have been converted in terms of $E_{sc}$ by using Eq. (2.2) – see the points in Fig. 8.

To draw the theoretical lines in Fig. 8, the parameters characterizing the $C_{10}E_5$, $C_{12}E_6$ and n-dodecanol molecules have been determined as explained in Sections 5.1 and 5.2 for $C_nE_m$ molecules; the only exception is that for dodecanol the value $a_{0,1} = 0.207$ nm$^2$ from Ref. [66] was used.

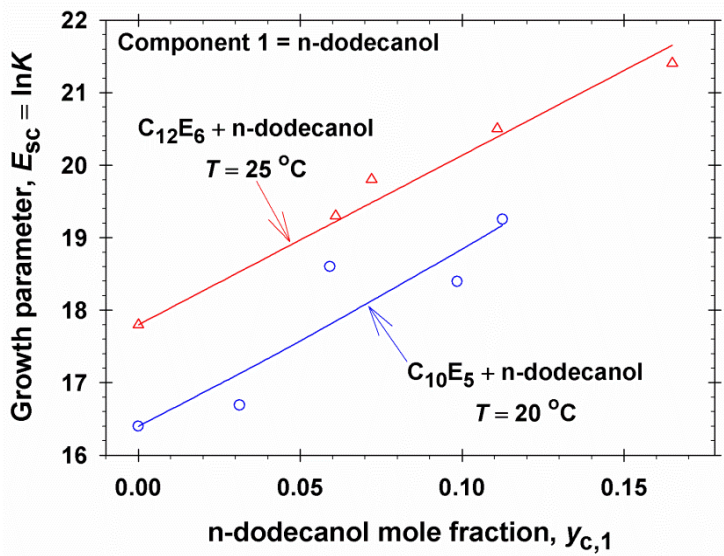

**Fig. 8.** Comparison of theory (solid lines) and experiment (points) for mixed micelles of $C_{10}E_5$ + n-dodecanol and $C_{12}E_6$ + n-dodecanol. Plots of the micelle growth parameter (scission energy in $k_BT$ units), $E_{sc}$, vs. the mole fraction of n-dodecanol, $y_{c,1}$.

As seen in Fig. 8, the experimental data are somewhat scattered, but the theoretical curves follow very well their tendency. Again, the theoretical curves have been drawn without using any adjustable parameters, and their agreement with the experimental data confirms the adequacy and reliability of the developed theoretical model.

## 8. Conclusions

In the present study, a quantitative molecular-thermodynamic theory of the growth of giant wormlike micelles of nonionic surfactants is developed on the basis of a generalized model, which includes the classical "phase separation" and "mass action" models [1-5] as special cases. The generalized model describes spherocylindrical micelles, which are simultaneously *multicomponent* and *polydisperse* in size. This model takes into account the fact that (in general) the micelle endcaps have a chemical composition, which is different from that of the cylindrical part of the micelle (Sections 3 and 4).

The molecular part of the model is based on explicit analytical expressions for the four components of the free energy of nonionic micelles: interfacial-tension, headgroup-steric, chain-conformation and free energy of mixing (Section 5). The radii of the cylindrical part and the spherical endcaps, $R_c$ and $R_s$, as well as the chemical composition of the endcaps, are determined by minimization of the free energy (Section 6).

A key new finding is that in the case of multicomponent micelles an additional term exists in the expression for the micelle growth parameter (micelle scission free energy), $E_{sc}$; see Eqs. (4.12) and (4.21). This term takes into account the fact that the micelle endcaps and cylindrical part have different compositions. The existence of this term has two important physical consequences: (i) It guarantees that the endcaps and the cylindrical part can coexist in chemical equilibrium with respect to exchange of all surfactant components; see Eqs. (6.2) and (6.3). (ii) Thanks to this term, the standard chemical potentials, which take into account contributions from internal degrees of freedom of the surfactant molecules, disappear from the final expression for $E_{sc}$, as it should be; see Eq. (4.24).

The theoretical model is tested against two sets of experimental data for wormlike micelles from binary surfactant mixtures: (i) surfactants with different headgroups but identical chains ($C_{14}E_5$ and $C_{14}E_7$) and (ii) surfactants with identical headgroups but different chains ($C_{14}E_5$ and $C_{10}E_5$); see Section 2. For both systems, excellent agreement between theory and experiment was achieved with respect to the experimental and theoretical values of $E_{sc}$ (and micelle mean mass aggregation number $n_M$) without using any adjustable parameters (Fig. 6). Good agreement between theory and experiment was achieved also for the mixed wormlike micelles from $C_{10}E_5$ and $C_{12}E_6$ with n-dodecanol (Fig. 8). In fact, the present article represents the first molecular thermodynamic study on the growth of mixed wormlike micelles, in which complete quantitative agreement between theory and experiment is achieved with respect to the prediction of micelle size (characterized by $n_M$). This is a considerable improvement over preceding studies [42-50].

For all investigated experimental systems, the calculated free energy possesses a minimum, which guarantees that the micelle exists in a state of mechanical and chemical equilibrium, as it should be for a physically adequate theory (Figs. 4 and 5). In addition to $E_{sc}$, the theory predicts the values of other micellar parameters, such as the radii of the cylindrical part and the endcaps, $R_c$ and $R_s$, as well as aggregation number and composition of the endcaps, $n_s$ and ($y_{s,1}, y_{s,2}$). It is difficult to directly measure these parameters, but information about their values and variations is provided by the theory (Fig. 7).

Another advantage of the molecular thermodynamic theory is that the derived analytical expressions for all basic micellar parameters allow their calculation by a standard personal computer or laptop. The fact that the mass averaged aggregation number of the wormlike micelles can be greater than $10^4$ monomers does not create any problems for the application of the developed analytical theory, whereas it could be an obstacle for the use of computer simulation methods like those in Refs. [71-73]. Appropriate combination of analytical and simulation methods could provide a fruitful way toward theoretical modelling of the growth of giant self-assembled molecular aggregates.

In future studies, the present molecular-thermodynamic approach can be extended to ionic and zwitterionic surfactants and their mixtures, which include amphiphilic molecules, fragrances and preservatives that are contained in typical formulations in personal-care and house-hold detergency.

**Acknowledgements**

The authors gratefully acknowledge the support from Unilever R&D, project No. MA-2018-00881N, and from the Operational Programme ''Science and Education for Smart Growth", Bulgaria, project No. BG05M2OP001-1.002-0012.

**Appendix A. Supplementary material**

(Attached below)

# Appendix A. Supplementary material

for the article

Analytical modeling of micelle growth. 2. Molecular thermodynamics of mixed aggregates and scission energy in wormlike micelles

<u>Authors</u>: Krassimir D. Danov, Peter A. Kralchevsky, Simeon D. Stoyanov, Joanne L. Cook, and Ian P. Stott

The cited equations and references have the same numbers as in the main text of the article. The notations have the same meaning as in the main text.

*A.1. Procedure for determination of $R_c$ and $R_s$ by minimization of $f_{c,int}$ and $E_{sc}$*

The equilibrium radius of the *cylindrical part* of a long ($n_c \gg n_s$) spherocylindrical micelle, $R_c$, has to be determined by minimization of the respective free energy per molecule, $f_c(\mathbf{y}_c, R_c)$, at fixed composition $\mathbf{y}_c = (y_{c,1}, y_{c,2}, ..., y_{c,m})$, which is determined by the input concentrations of the surface active species. Consequently, in view of Eqs. (4.6), (4.22) and (5.1), the equilibrium value of $R_c$ has to be determined from the equation:

$$\frac{\partial f_{c,mix}}{\partial R_c} + \frac{\partial f_{c,\sigma}}{\partial R_c} + \frac{\partial f_{c,hs}}{\partial R_c} + \frac{\partial f_{c,conf}}{\partial R_c} = 0 \tag{A.1}$$

Eq. (A.1) represents an implicit equation for $R_c$, which has been solved numerically. To achieve the best accuracy of the numerical solution, the derivatives have been calculated analytically (see below) and the obtained expressions have been substituted in Eq. (A.1).

The equilibrium radius, $R_s$, and the composition, $\mathbf{y}_s = (y_{s,1}, y_{s,2}, ..., y_{s,m})$, of the *spherical endcaps* of a long spherocylindrical micelle have to be determined by numerical minimization of the excess free energy per molecule in the endcaps, $E_{sc}$, at fixed composition, $\mathbf{y}_c = (y_{c,1}, y_{c,2}, ..., y_{c,m})$, and radius, $R_c$, of the cylindrical part. The expression for $E_{sc}$,

$$E_{sc} k_B T = n_s [f_{s,int}(R_s, \mathbf{y}_s, R_c) - f_{c,int}(\mathbf{y}_c, R_c)] - n_s \sum_{j=1}^{m} (y_{s,j} - y_{c,j}) \frac{\partial f_{c,int}(\mathbf{y}_c, R_c)}{\partial y_{c,j}} \tag{A.2}$$

contains the derivative of $f_{c,int}$ with respect to the endcap composition $y_{c,j}$; see Eq. (4.24). Again, to achieve the best accuracy of the numerical minimization of $E_{sc}$, these derivatives have been calculated analytically (see below) and the obtained expressions have been substituted in Eq. (A.2).

*A.2. Derivatives of the free energy of mixing $f_{c,mix}$*

In this article, the molecular thermodynamic theory is verified against experimental data for *binary* mixtures of nonionic surfactants. In view of Eqs. (5.15) and (5.16), for a binary surfactant mixture $f_{c,mix}$ can be presented in the form:

$$\frac{f_{c,mix}}{k_B T} = y_{c,1} \ln(\frac{y_{c,1} v_1}{y_{c,1} v_1 + y_{c,2} v_2}) + y_{c,2} \ln(\frac{y_{c,2} v_2}{y_{c,1} v_1 + y_{c,2} v_2}) \tag{A.3}$$

Our first goal is to calculate the derivatives of $f_{c,mix}$ with respect to $R_c$, $y_{c,1}$ and $y_{c,2}$. In view of Eq. (A.3), $f_{c,mix}$ is independent of $R_c$, so that

$$\frac{\partial f_{c,mix}}{\partial R_c} = 0 \tag{A.4}$$

The other two derivatives are:

$$\frac{\partial f_{c,mix}}{\partial y_{c,1}} = k_B T [\ln(\frac{y_{c,1} v_1}{y_{c,1} v_1 + y_{c,2} v_2}) + 1 - \frac{v_1}{y_{c,1} v_1 + y_{c,2} v_2}] \tag{A.5}$$

$$\frac{\partial f_{c,mix}}{\partial y_{c,2}} = k_B T [\ln(\frac{y_{c,2} v_2}{y_{c,1} v_1 + y_{c,2} v_2}) + 1 - \frac{v_2}{y_{c,1} v_1 + y_{c,2} v_2}] \tag{A.6}$$

*A.3. Derivatives of the interfacial tension component $f_{c,\sigma}$*

In view of Eqs. (5.10), (5.18) and (5.19), for a binary surfactant mixture Eqs. (5.17), (5.20) and (5.21) can be presented in the form:

$$f_{c,\sigma} = \frac{\sigma_{ow} R_c}{R_c + \delta_T}[\frac{2}{R_c}(y_{c,1} v_1 + y_{c,2} v_2) - (y_{c,1} a_{0,1} + y_{c,2} a_{0,2})] \tag{A.7}$$

$$\sigma_{ow} = 47.12 + 1.479(y_{c,1} n_1 + y_{c,2} n_2)^{0.5422} - 0.0875(T - 293) \tag{A.8}$$

$$\delta_T = 0.1456(y_{c,1} l_1 + y_{c,2} l_2) \tag{A.9}$$

The partial derivative of Eq. (A.7) with respect to $R_c$ reads:

$$\frac{\partial f_{c,\sigma}}{\partial R_c} = \frac{\delta_T f_{c,\sigma}}{R_c(R_c + \delta_T)} - \frac{\sigma a_c}{R_c} \tag{A.10}$$

where $f_{c,\sigma}$, $\sigma$ and $\delta_T$ are given by Eqs. (5.17)–(5.21). In addition, the expressions for the partial derivatives of $f_{c,\sigma}$ with respect to the molar fractions are:

$$\frac{\partial f_{c,\sigma}}{\partial y_{c,1}} = \frac{0.8019 n_1 f_{c,\sigma}}{\sigma_{ow}(y_{c,1}n_1 + y_{c,2}n_2)^{0.4578}} - \frac{0.1456 l_1 f_{c,\sigma}}{R_c + \delta_T} + \frac{\sigma_{ow} R_c}{R_c + \delta_T}(\frac{2}{R_c} v_1 - a_{0,1}) \quad (A.11)$$

$$\frac{\partial f_{c,\sigma}}{\partial y_{c,2}} = \frac{0.8019 n_2 f_{c,\sigma}}{\sigma_{ow}(y_{c,1}n_1 + y_{c,2}n_2)^{0.4578}} - \frac{0.1456 l_2 f_{c,\sigma}}{R_c + \delta_T} + \frac{\sigma_{ow} R_c}{R_c + \delta_T}(\frac{2}{R_c} v_2 - a_{0,2}) \quad (A.12)$$

*A.4. Derivatives of the headgroup steric repulsion component $f_{c,hs}$*

Substituting Eqs. (5.10), (5.23) and (5.24) into Eq. (5.22), we obtain:

$$f_{c,hs} = -k_B T \ln[1 - \frac{R_c(y_{c,1}^2 a_{0,1} + 2 y_{c,1} y_{c,2} a_{12} + y_{c,2}^2 a_{0,2})}{2(y_{c,1}v_1 + y_{c,2}v_2)}] \quad (A.13)$$

where $a_{12}$ is defined by the expression:

$$a_{12} = \frac{1}{4}(a_{0,1} + 2\sqrt{a_{0,1} a_{0,2}} + a_{0,2}) \quad (A.14)$$

see Eq. (5.25). The partial derivative of $f_{hs}$ with respect to $R_c$ is:

$$\frac{\partial f_{c,hs}}{\partial R_c} = \frac{k_B T}{R_c} \frac{a_{hs}}{a_c - a_{hs}} \quad (A.15)$$

where $a_c$ and $a_{hs}$ are given by Eqs. (5.10) and (5.23). The differentiation of Eq. (A.13) with respect to $y_{c,1}$ and $y_{c,2}$, after some transformations yields:

$$\frac{\partial f_{c,hs}}{\partial y_{c,1}} = 2k_B T \frac{y_{c,1}a_{0,1} + y_{c,2}a_{1,2}}{a_c - a_{hs}} - 2k_B T \frac{v_1}{R_c a_c} \frac{a_{hs}}{a_c - a_{hs}} \quad (A.16)$$

$$\frac{\partial f_{c,hs}}{\partial y_{c,2}} = 2k_B T \frac{y_{c,2}a_{0,2} + y_{c,1}a_{1,2}}{a_c - a_{hs}} - 2k_B T \frac{v_2}{R_c a_c} \frac{a_{hs}}{a_c - a_{hs}} \quad (A.17)$$

*A.5. Derivatives of the chain conformation component $f_{c,conf}$*

For the cylindrical part of the micelle, $p_c = 1/2$ and $c_{conf} = 1/3$. Then, Eq. (5.28) acquires the form:

$$\frac{f_{c,\text{conf}}}{k_B T} = \frac{\pi^2 R_c^2}{16 l_{sg}} \left[ \frac{y_{c,1}}{l_1} + \frac{y_{c,2}}{l_2} - (y_{c,1} l_1 + y_{c,2} l_2)(\frac{1}{l_2^2} - \frac{1}{l_1^2}) \beta_{\text{conf}} \right] \quad (A.18)$$

where

$$\beta_{\text{conf}}(\eta_{c,1}) = (1-b^2)^{3/2} + \frac{3}{2} b^2 \eta_{c,1} - \eta_{c,1} \quad (A.19)$$

$$\eta_{c,1} = \frac{y_{c,1} v_1}{y_{c,1} v_1 + y_{c,2} v_2} \quad (A.20)$$

and $b(\eta_{c,1})$ is the solution of the transcendent equation

$$(1-b^2)^{1/2} - b^2 \ln\left[\frac{1}{b} + (\frac{1}{b^2} - 1)^{1/2}\right] = \eta_{c,1} \quad (A.21)$$

see Eqs. (5.16), (5.31) and (5.32). First, from Eq. (A.18) we get:

$$\frac{\partial f_{c,\text{conf}}}{\partial R_c} = \frac{2}{R_c} f_{c,\text{conf}} \quad (A.22)$$

The partial derivatives with respect to the molar fractions are:

$$\frac{\partial f_{c,\text{conf}}}{\partial y_{c,1}} = k_B T \frac{\pi^2 R_c^2}{16 l_{sg} l_1} \left[1 - (\frac{l_1^2}{l_2^2} - 1)\beta_{\text{conf}}(\eta_{c,1}) - (y_{c,1} + y_{c,2}\frac{l_2}{l_1})(\frac{l_1^2}{l_2^2} - 1)\frac{\partial \beta_{\text{conf}}}{\partial y_{c,1}}\right] \quad (A.23)$$

$$\frac{\partial f_{c,\text{conf}}}{\partial y_{c,2}} = k_B T \frac{\pi^2 R_c^2}{16 l_{sg} l_2} \left[1 - (1 - \frac{l_2^2}{l_1^2})\beta_{\text{conf}}(\eta_{c,1}) - (y_{c,1}\frac{l_1}{l_2} + y_{c,2})(1 - \frac{l_2^2}{l_1^2})\frac{\partial \beta_{\text{conf}}}{\partial y_{c,2}}\right] \quad (A.24)$$

The partial derivatives of the interaction parameter $\beta_{\text{conf}}$, which appear in the right-hand sides of Eqs. (A.23) and (A.24), can be calculated as follows. First we calculate the derivative of Eq. (A.21) with respect to $\eta_{c,1}$:

$$2b \ln\left[\frac{1}{b} + (\frac{1}{b^2} - 1)^{1/2}\right] \frac{db}{d\eta_{c,1}} = -1 \quad (A.25)$$

Next, the logarithmic term is eliminated between Eqs. (A.21) and (A.25):

$$\frac{2}{b}[\eta_{c,1} - (1-b^2)^{1/2}] \frac{db}{d\eta_{c,1}} = 1 \quad (A.26)$$

Further, we calculate the derivative of Eq. (A.19) with respect to $\eta_{c,1}$:

$$\frac{d\beta_{\text{conf}}}{d\eta_{c,1}} = 3b[\eta_{c,1} - (1-b^2)^{1/2}] \frac{db}{d\eta_{c,1}} + \frac{3}{2} b^2 - 1 \quad (A.27)$$

The combination of Eqs. (A.26) and (A.27) leads to the following simpler expression:

$$\frac{d\beta_{conf}}{d\eta_{c,1}} = 3b^2 - 1 \tag{A.28}$$

Finaly, in view of Eqs. (A.20) and (A.28) we obtain:

$$\frac{\partial \beta_{conf}}{\partial y_{c,1}} = \frac{d\beta_{conf}}{d\eta_{c,1}} \frac{\partial \eta_{c,1}}{\partial y_{c,1}} = (3b^2 - 1)\frac{v_1 v_2 y_{c,2}}{(y_{c,1}v_1 + y_{c,2}v_2)^2} \tag{A.29}$$

$$\frac{\partial \beta_{conf}}{\partial y_{c,2}} = \frac{d\beta_{conf}}{d\eta_{c,1}} \frac{\partial \eta_{c,1}}{\partial y_{c,2}} = -(3b^2 - 1)\frac{v_1 v_2 y_{c,1}}{(y_{c,1}v_1 + y_{c,2}v_2)^2} \tag{A.30}$$

Thus, the derivatives of $f_{c,conf}$ with respect to the molar fractions $y_{c,1}$ and $y_{c,2}$ are to be calculated from Eqs. (A.23) and (A.24), along with Eqs. (A.29) and (A.30).